\documentclass[aps,prd,reprint,superscriptaddress]{revtex4-1}

\pdfoutput=1

\usepackage{upgreek}

\usepackage{mathtools}
\usepackage{amssymb}
\usepackage{amsfonts}
\usepackage{mathrsfs}

\usepackage{graphicx}
\graphicspath{{./}}
\usepackage{url}
\usepackage{hyperref}

\usepackage{siunitx}
\usepackage{enumitem}
\makeatletter
\let\orig@Itemize =\itemize
\let\orig@Enumerate =\enumerate
\let\orig@Description =\description
\renewenvironment{itemize}{\orig@Itemize[noitemsep,nolistsep]}{\endlist}
\renewenvironment{enumerate}{\orig@Enumerate[noitemsep,nolistsep]}{\endlist}
\renewenvironment{description}{\orig@Description[noitemsep,nolistsep]}{\endlist}
\makeatother

\usepackage{accents}

\newcommand*{\ud}{\mathrm{d}}
\newcommand*{\s}{\,}
\newcommand*{\anu}{\ensuremath{\bar{\nu}_e}}
\newcommand*{\nue}{\ensuremath{\nu_e}}
\newcommand*{\Dmn}{\ensuremath{\Delta m^2_\mathrm{new}}}
\newcommand*{\stn}{\ensuremath{\sin^2(2\theta_\mathrm{new})}}
\newcommand*{\tn}{\ensuremath{\theta_\mathrm{new}}}

\newcommand*{\ce}{$^{144}\text{Ce}$}
\newcommand*{\pr}{$^{144}\text{Pr}$}
\newcommand*{\nd}{$^{144}\text{Nd}$}
\newcommand*{\ma}{FSUE ``Mayak'' PA}
\newcommand{\nuc}[3]{$\prescript{#2}{#3}{\text{#1}}$}

\newcommand{\Ar}[1]{$^{#1}\text{Ar}$}
\newcommand*{\K}{$^{42}\text{K}$}
\newcommand*{\Cr}{$^{51}\text{Cr}$}
\newcommand*{\Ru}{$^{106}\text{Ru}$}
\newcommand*{\Rh}{$^{106}\text{Rh}$}
\newcommand*{\Sr}{$^{90}\text{Sr}$}
\newcommand*{\Y}{$^{90}\text{Y}$}

\newcommand*{\uc}{$^{235}\text{U}$}
\newcommand*{\uh}{$^{238}\text{U}$}
\newcommand*{\pu}{$^{239}\text{Pu}$}

\newcommand*{\cm}{$^{244}\text{Cm}$}

\newcommand*{\Eres}{\SI{7.5}{\%}}
\newcommand*{\Lres}{\SI{15}{cm}}

\begin{document}

\title{Experimental Parameters for a Cerium 144 Based Intense Electron\\ 
Antineutrino Generator Experiment at Very Short Baselines}

\author{J.~Gaffiot}
\email{jonathan.gaffiot@cea.fr}
  \affiliation{Astroparticules et Cosmologie APC, 10 rue Alice Domon et L\'eonie Duquet,
    75205 Paris cedex 13, France}
\author{T.~Lasserre}
  \affiliation{Astroparticules et Cosmologie APC, 10 rue Alice Domon et L\'eonie Duquet,
    75205 Paris cedex 13, France}
  \affiliation{Commissariat a l'\'energie atomique et aux \'energies alternatives,
    Centre de Saclay, IRFU, 91191 Gif-sur-Yvette, France}
\author{G.~Mention}
  \affiliation{Commissariat a l'\'energie atomique et aux \'energies alternatives,
    Centre de Saclay, IRFU, 91191 Gif-sur-Yvette, France}
\author{M.~Vivier}
  \affiliation{Commissariat a l'\'energie atomique et aux \'energies alternatives,
    Centre de Saclay, IRFU, 91191 Gif-sur-Yvette, France}
\author{M.~Cribier}
  \affiliation{Astroparticules et Cosmologie APC, 10 rue Alice Domon et L\'eonie Duquet,
    75205 Paris cedex 13, France}
  \affiliation{Commissariat a l'\'energie atomique et aux \'energies alternatives,
    Centre de Saclay, IRFU, 91191 Gif-sur-Yvette, France}
\author{M.~Durero}
\author{V.~Fischer}
\author{A.~Letourneau}
  \affiliation{Commissariat a l'\'energie atomique et aux \'energies alternatives,
    Centre de Saclay, IRFU, 91191 Gif-sur-Yvette, France}
\author{E.~Dumonteil}
  \affiliation{Commissariat a l'\'energie atomique et aux \'energies alternatives,
    Centre de Saclay, SERMA, 91191 Gif-sur-Yvette, France}
\author{I.~S.~Saldikov}
\author{G.~V.~Tikhomirov}
  \affiliation{National Research Nuclear University MEPhI (Moscow Engineering Physics Institute)}
\date{\today}

\begin{abstract}
The standard three-neutrino oscillation paradigm, associated with small squared mass splittings $\Delta m^2
\ll \SI{0.1}{eV^2}$, has been successfully built up over the last \SI{15}{years} using solar, atmospheric,
long baseline accelerator and reactor neutrino experiments.  However, this well-established picture might
suffer from anomalous results reported at very short baselines in some of these experiments. If not
experimental artifacts, such results could possibly be interpreted as the existence of at least an additional
fourth sterile neutrino species, mixing with the known active flavors with an associated mass splitting
$\Dmn \gg \SI{0.01}{eV^2}$, and being insensitive to standard weak interactions. Precision
measurements at very short baselines (\SIrange{5}{15}{m}) with intense \si{MeV} \anu{} emitters can be used to
probe these anomalies. In this article, the expected \anu{} signal and backgrounds of a generic experiment which consists of
deploying an intense $\upbeta^-$ radioactive source inside or in  the vicinity of a large liquid scintillator
detector are studied. The technical challenges to perform such an experiment are identified, along with quantifying the
possible source and detector induced systematics, and their impact on the sensitivity to the observation
of neutrino oscillations at short baselines.
\end{abstract}

\maketitle

\section{Introduction}
Neutrino oscillations have been observed in solar, atmospheric, long baseline accelerator and reactor neutrino
experiments~\cite{Beringer:1900zz}. The data collected so far by these experiments are well described in the framework of
a three active neutrino mixing approach, in which the three known flavor neutrinos ($\nu_e, \nu_\mu, \nu_\tau$), are
unitary linear combinations of three mass states ($\nu_1, \nu_2, \nu_3$), with squared mass differences of
$\Delta m_{21}^2 = \Delta m_{\mathrm{sol}}^2 = 7.50_{-0.20}^{+0.19} \times \SI{e-5}{eV^2}$ and
$\mid\Delta m_{31}^{2}\mid \approx \mid\Delta m_{32}^{2}\mid = \Delta m_{\mathrm{atm}}^2=2.32_{-0.08}^{+0.12}
\times \SI{e-3}{eV^2}$~\cite{Beringer:1900zz}. The ``sol'' and ``atm'' subscripts stand historically for solar
and atmospheric experiments, which provided the first evidences for neutrino oscillations.
Beyond this well-established picture, anomalous results have been reported in different neutrinos experiments,
such as the LSND~\cite{Aguilar:2001ty} and MiniBooNE~\cite{AguilarArevalo:2008rc,AguilarArevalo:2010wv}
accelerator experiments, the calibration of the Gallex and SAGE solar neutrino
experiments~\cite{Giunti:2006bj,Giunti:2010zu,Giunti:2012tn} and, more recently, the reactor experiments at short
baselines~\cite{Mention:2011}. If not related to yet non-understood experimental issues, global fits of
short-baseline neutrino oscillation experiments show that these anomalies can be interpreted by the addition
of one (3+1) or two (3+2) sterile neutrinos to the standard three-neutrino paradigm, although no sterile
neutrino model currently provides a compelling explanation of all data~\cite{Mention:2011}.

This article addresses the possibility and technical feasibility of deploying an intense \anu{} generator,
using a radioactive $\upbeta^-$ decaying isotope, outside or inside a large liquid scintillator detector to
test these short baseline anomalies. The key experimental parameters of such a radioactive source experiment
are identified, and their impact on the sensitivity to short baseline neutrino oscillation is studied. The
simplest (3+1) neutrino mixing scheme is considered here, with a fourth massive neutrino state presenting
squared mass splittings in the range $\lvert \Dmn \rvert \sim \SIrange[range-phrase = -]{0.1}{5}{eV^2}$.

The article is organized as follows: section II describes the motivations of such an experimental concept,
while section III details the choice of the radioactive source to be used as an \anu{} generator. Section IV
reports on source-induced backgrounds, activity and \anu{} energy spectra, which must be accurately
characterized in order not to degrade the experimental sensitivity to short baseline oscillations.
Section V reviews the existing large liquid scintillator detectors, which meet our experimental requirements.
Sections VI and VII describe the signature of neutrino oscillations in such large liquid scintillator
detectors, as well as the impact of source-related and detector-related experimental parameters on the
sensitivity to short baseline oscillations. As explained in section IV, minimizing the source-induced
backgrounds is of primary importance: section VIII therefore presents the results of an accurate simulation
of the source-induced backgrounds in a large liquid scintillator detector to set the source maximum levels of
radioactive impurities. Finally, section IX is devoted to our conclusions.

\section{Experimental Concept}
\label{sec:exp_concept}
An experiment must be sensitive to the deformation of the \anu{} energy spectrum, and/or the deformation of
the spatial distribution of  \anu{} interactions to unambiguously test the short baseline oscillation
hypothesis. The $\Delta m^2$ region coverage is mostly driven by the source-detector distance. Using \anu{}
with MeV energies, distances of the order of 1 to \SI{10}{m} are required to probe the \SIrange{0.1}{5}{eV^2}
$\Delta m^2$ region preferred by global fits to short baseline neutrino data. A solution meeting this
source-detector requirement has already been proposed by placing a ``small'' liquid scintillator based \anu{}
detector close to a nuclear reactor~\cite{white}. This article studies an alternative solution, consisting in
deploying an \anu{} generator in the vicinity of a liquid scintillator detector, as already proposed
in~\cite{Ianni:1999nk,Cribier:2011fv}. Such a solution requires a high-enough source activity and a large-enough detector
in order to accumulate a significant statistics over a few mean life times of the $\upbeta^-$ decaying
radioisotope. Furthermore, a good detector must have the capability of precisely reconstructing both the
energy and interaction vertex of each neutrino candidate, in order to observe an oscillation pattern in the
$L/E\sim \SIrange[range-phrase = -]{0.1}{10}{m\,MeV^{-1}}$ range. Three suitable detectors, which are (or are
soon-to-be) in operation, have been identified: KamLAND~\cite{Eguchi:2002dm}, Borexino~\cite{Alimonti:2008gc}
and SNO+~\cite{Chen:2008un}.

They all share features that are important to a clean and unambiguous measurement: they are located deep
underground in an ultra low background environment,  they are $\cal{O}$(\SI{10}{m})-sized spherical detector
filled
up with ultra pure hydrogen-rich liquid scintillator and have good vertex reconstruction and energy
resolution.

A statistics of a few \num{10000} of \anu{} interactions, collected over a few emitter's mean life times, is
necessary to probe the $\Delta m^2$ region preferred by the global fits to short baseline anomalous results.
Considering the typical dimensions of the detectors mentioned above, source activities must range from
\SI{1}{PBq} to \SI{10}{PBq}~\footnote{\SI{1}{Ci} = \SI{3.7e10}{Bq}, \SI{1}{PBq} = \SI{27}{kCi}}, depending on
the source-detector distance and detection volume (fiducial mass of liquid scintillator).

\section{Source radioisotope}
The choice of using a \nue{} or an \anu{} source is made upon the consideration of many factors, such as for
example source mean life time, source production feasibility, signal detection process and backgrounds either
coming from the source or the detector environment. This section reviews the advantages and drawbacks of  such
\nue{} and \anu{} sources. A focus to the \ce{}-\pr{} pair of radioisotopes, which is to our opinion the best
candidate as a radioactive source to be deployed close to a large liquid scintillator detector, is presented.

\subsection{Neutrino sources}
Production of intense \nue{} sources and deployment in neutrino detectors have already been performed in the
90s, for the precise understanding of the Gallex and SAGE~\cite{Cribier:1996cq,Abdurashitov:1998ne,Abdurashitov:2005tb}
radiochemical solar neutrino detectors. Such sources emit \nue{} through electron capture process of
unstable nuclei. Two nuclei were used at that time, and are still under consideration for a radioactive source
experiment at short baselines: \Cr{} and \Ar{37}.

\Cr{} decays with a \SI{27.7}{days} half-life, and mostly produces \SI{753}{keV} neutrinos. However,
\SI{433}{keV} neutrinos are produced with a \SI{10}{\%} branching ratio along with a \SI{320}{keV} gamma.
\Ar{37} produces \SI{814}{keV} neutrinos with \SI{100}{\%} branching ratio and a  \SI{35}{days}
half-life~\cite{NNDC}. In terms of heat release, and shielding to gamma rays, \Ar{37} is easier than \Cr{} to
handle. It also benefits from a slightly longer half-life and a higher neutrino energy which could help
discriminating against gammas from natural radioactivity. Being a metal \Cr{} is chemically easier
to extract and manipulate than \Ar{37} .

The realization of such \nue{} sources is technically challenging. Both isotopes have to be produced by neutron
irradiation inside a nuclear reactor, through \nuc{Cr}{50}{}(n,$\upgamma$)\Cr{} and
\nuc{Ca}{40}{}(n,$\upalpha$)\Ar{37} reactions for \Cr{} and \Ar{37}, respectively. A very high neutron flux is
necessary with multiple irradiation steps to achieve a high-enough specific activity. Furthermore, the
(n,$\upalpha$) reaction has a \SI{1.75}{MeV} threshold requiring irradiation with fast neutrons to efficiently
produce the required level of activity. This is an additional technical challenge to the production of a
\Ar{37} source.

From the transportation point of view, the deployment of \Cr{} or \Ar{37} source is also  challenging. As
pointed out above, the half-lives of such radionuclides are of the order of a few tens of days, requiring
prompt transportation and deployment logistic solutions to be implemented. 
This could be a major issue when the suitable nuclear reactor for neutron irradiation and the deployment site
are far away, and especially if they are located in two different countries. Moreover, nuclear
safety regulations concerning the transportation of radioactive materials are very strict and can for example
forbid air transportation schemes.

Last but not least, \nue{} are detected through inelastic scatterings off electrons, $\nue{} + e^- \rightarrow
\nue{} + e^-$, having a small cross-section relative to the inverse $\upbeta$ decay process often used to
detect $\anu{}$. Moreover, the detection is very sensitive to backgrounds because the scattering of 
$\cal{O}(\SI{1}{MeV})$ \nue{} lies in the natural $\upbeta$ or $\upgamma$ ray radioactivity energy range. So 
far, only the Borexino detector, which is designed to study low energy solar neutrinos, has demonstrated a 
low-enough background level to efficiently separate a low-rate electron scattering $\nue{}$ signal from
backgrounds~\cite{Alimonti:2008gc}. Finally, an activity of \SI{\sim500}{PBq} is necessary to correctly probe
the short baseline anomalies.

\subsection{Antineutrino sources}
As opposed to a \nue{} source, an \anu{} source, or AntiNeutrino Generator (ANG hereafter), is a $\upbeta^-$
decaying nucleus producing \anu{} over a broad energy spectrum, up to the maximum endpoint energy ($\sim
Q_\upbeta$) of its available $\upbeta$ branches. The detection of \anu{} in liquid scintillator detectors
relies on the Inverse Beta Decay (IBD) reaction: $\anu{} + p \rightarrow e^+ + n$. The IBD reaction
cross-section is higher than the cross-section of neutrino scattering on electrons by roughly an order of
magnitude at MeV energies. Furthermore, the IBD reaction signature is a time and space coincidence between the
positron prompt energy deposition ($\ud E/\ud x$ + annihilation) and the delayed gamma energy deposition
coming from neutron capture. This signature allows a very efficient IBD candidate selection
together with a powerful background rejection. The prompt signal visible energy is $E^{\rm vis}_{\rm prompt} =
E_{\bar{\nu_e}}-Mc^2+2m_ec^2$, where $M=\SI{1.293}{MeV/c^2}$ the mass difference between proton and neutron
and $m_e=\SI{0.511}{MeV/c^2}$ the electron mass, so that $\SI{1.022}{MeV} < E^{\rm vis}_{\rm prompt} < 
Q_\upbeta - \SI{0.782}{MeV}$. In a non-doped scintillator, neutrons are mostly captured on hydrogen atoms, 
which then release a \SI{2.2}{MeV} gamma ray. The time distribution between the prompt and delayed events 
follows an exponential law with a time constant which is equal to the capture time of neutrons on hydrogen
$\tau\sim\SI{200}{\micro\s}$. Backgrounds to the IBD signal selection are of two types. The first type is the
accidental background, which is made from two random energy depositions in a time window
roughly corresponding to the hydrogen capture time. For example, such accidental coincidences arise from two
gammas coming from the surrounding natural radioactivity and interacting in the target. The second type of
background is called correlated background. For instance, spallation of cosmic rays produces fast neutrons,
that  can thermalize and also be captured in the liquid scintillator, faking both a prompt and delayed energy
deposition. Fortunately the above-mentioned backgrounds are negligible in the three detectors under
consideration thanks to large overburdens, very low radioactivity materials and thick shieldings from passive
materials.

The IBD reaction energy threshold is \SI{1.806}{MeV}, and requires a source radioisotope with a high endpoint
$\upbeta^-$ decay. Since half-life and endpoint energy are strongly anti-correlated quantities for $\upbeta^-$
decay, this requirement leads to look for nuclei with half-lives typically shorter than a day, then preventing
the production and use of an ANG made of a single radioisotope. However, looking for a cascading pair of
$\upbeta^-$ decaying isotopes, the parent nucleus being a rather long-lived isotope (with month or year-scale
half-life) and the daughter nucleus being a short-live isotope, could circumvent this difficulty. Another requirement
is that the daughter isotope must have a $\upbeta^-$ endpoint energy as high as possible above the IBD
threshold to maximize the IBD reaction rate. Several pairs of isotopes meeting these requirements have been
identified by browsing nuclear databases and are displayed in table~\ref{tab:nubar_gen_couples}. Among the
identified isotopes, the choice of the pair making the best ANG to test the short baseline anomalies is
detailed in the next section.

\begin{table}[!ht]
 \centering
 \begin{tabular}{c c c}
  \toprule
  	\rule{0pt}{2.25ex}
  	Couple & $\tau_{1/2}$ of parent & $Q_{\upbeta^-}$ of daughter \tabularnewline
\colrule
 	\rule{0pt}{2.5ex}
  	\Ar{42} - \K{}	& \SI{33}{y}	& \SI{3.53}{MeV} \tabularnewline
  	\Sr{} - \Y{}	& \SI{28.9}{y}	& \SI{2.28}{MeV} \tabularnewline
 	\Ru{} - \Rh{}	& \SI{372}{d}	& \SI{3.55}{MeV} \tabularnewline
  	\ce{} - \pr{}	& \SI{285}{d}	& \SI{3.00}{MeV} \tabularnewline
  \botrule
 \end{tabular}
 \caption{\label{tab:nubar_gen_couples} Suitable couple for an \anu{} radioactive
source~\cite{NNDC}.}
\end{table}

\subsection{The golden pair: cerium - praseodymium 144}
The choice of the best couple as a suitable ANG is also driven by production feasibility. \Sr{}, \Ru{} and
\ce{} isotopes are fission products found in nuclear reactors and could be extracted from nuclear spent fuel.
\Ar{42} has to be produced through fast neutron irradiation on \Ar{41}, which has a very short half-life
($T_{1/2}$ = \SI{109}{ms}). Production of \Ar{42} then requires a challenging double neutron capture starting
from \Ar{40}. It has therefore been rejected as a possible candidate.

Only \Sr{}, \Ru{} and \ce{} then remain as good candidates to make a suitable ANG. They now have to be
compared on the basis of their respective production rate in nuclear reactor cores. The cumulative fission
yield is the number of nuclei produced per fission when the reactor core is at equilibrium, particularly
including decays from short-lived fission products. To a first approximation (long-lived isotopes never reach
equilibrium during fuel irradiation), this quantity allows the comparison of the abundance of each couple.
It is shown in table~\ref{tab:cum_FY} for thermal fission of \uc{} and \pu{}, which are by far
the most abundant fissioned nuclei in current nuclear reactors. \Ru{} is disfavoured because of its low
fission yield for \uc{}, while \ce{} and \Sr{} have comparable fission yields for both \uc{} and \pu{}.

\begin{table}[!ht]
 \centering
 { \sisetup{separate-uncertainty=false}
 \begin{tabular}{c c c c}
  \toprule
  	\rule{0pt}{2.25ex} & \multicolumn{3}{c}{Cumulative fission yield (\si{\%})}
\tabularnewline
  	\rule{0pt}{2.5ex}  & \Ru{} & \Sr{} & \ce{} \tabularnewline \colrule
 	\rule{0pt}{2.5ex} \uc{} & \num{0.401(6)} & \num{5.78(6)} & \num{5.50 (4)}
\tabularnewline
 	\rule{0pt}{2.5ex} \pu{} & \num{4.35(9)} & \num{2.10(4)} & \num{3.74 (3)}
\tabularnewline
  \botrule
 \end{tabular}
 }
 \caption{\label{tab:cum_FY} Cumulative thermal fission yields of \Ru{}, \Sr{} and
\ce{}~\cite{NNDC} for the two main fissile isotopes fuelling nuclear reactors.}
\end{table}

Maximizing both the endpoint energy of the daughter nucleus and the source activity (driven by the decay
period of the parent nucleus) ultimately leads to reject the \Sr{} - \Y{} couple. Then, \ce{} - \pr{} remains
as the best couple for an ANG. Finally, the chemical separation of cerium from other lanthanides is feasible
at the industrial scale (see~\ref{sec:compo}).

\begin{figure*}[ht]
\centering \includegraphics[width=0.8\linewidth]{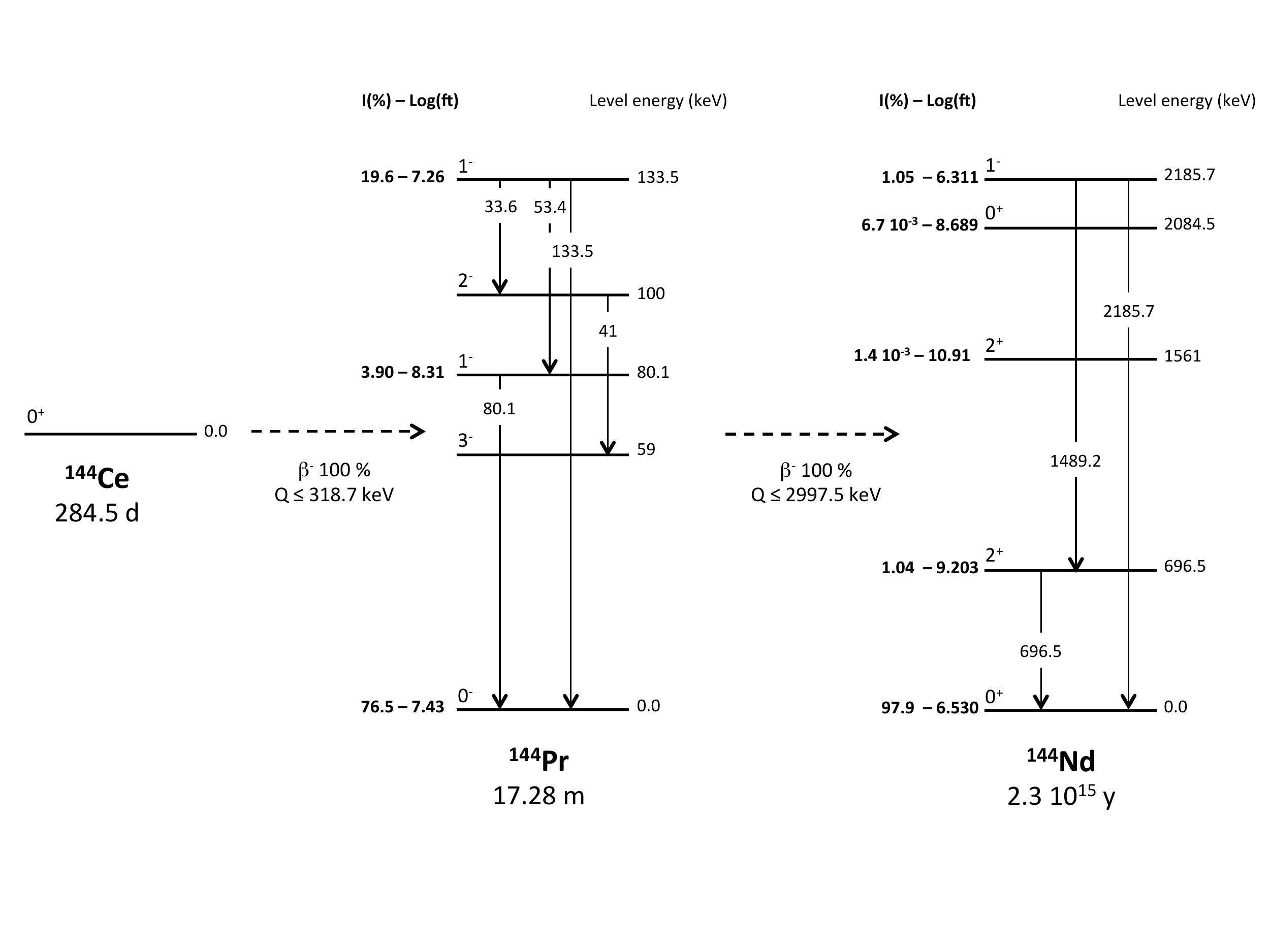}
\caption{\label{fig:144Ce-144Pr} Simplified decay scheme of the \ce{}-\pr{}
couple. $\upbeta$ branches with branching ratios greater than \SI{0.001}{\%} are displayed, along with the
corresponding Log(ft) values, daughter nucleus level energies and spin parities. The main gamma transitions
(intensity greater than \SI{0.1}{\%}) among the excited states of the \pr{} and \nd{} nuclei are also
displayed together with their corresponding energies. The full \ce{} and \pr{} decay scheme data can be found
in~\cite{NNDC}.}
\end{figure*}

The most suitable ANG for a very short baseline experiment is therefore the \ce{}-\pr{} pair.
Depending on the source-detector distance and on the detector target mass, an activity of \SI{2}{PBq} to
\SI{10}{PBq} (relative to \ce{} $\upbeta^-$ decay rate) is necessary to achieve a statistics of 10 000 IBD
candidates. This is a factor 10 below the required activities for \nue{} sources in the same experimental
configuration.

However, \pr{} $\upbeta^-$ decay is followed \SI{0.7}{\%}  of the time by a \SI{2.185}{MeV} gamma ray, as can
be seen on the \ce{}-\pr{} decay scheme presented on figure~\ref{fig:144Ce-144Pr}.
This gamma ray could fall both in the prompt and delayed energy windows, then being an additional source of
accidental background. Moreover, due to the very high activity of the ANG, this ray constitutes a major
radiation protection concern. A dedicated high-Z material shielding, made for instance of lead or tungsten, is
therefore necessary to suppress this background. As a reference, a \SI{19}{cm} thick tungsten alloy shielding
with typical density of \SI{18}{g/cm^3} provides an attenuation of \num{3.1e-7} to the \SI{2.185}{MeV} gamma
ray, corresponding to a dose at \SI{1}{m} of \SI{0.01}{mSv/h} for a \SI{5}{PBq} source.
Finally, it is worth noting that the shielding thickness shall also be designed to comply with dose limits
imposed by any national, institutional or international regulation.

\section{Source characterization}
Characterizing the \ce{}-\pr{} source is a key requirement for achieving a good experimental sensitivity to
short baseline neutrino oscillations. First, source composition and level of radioactive impurities have to be
accurately controlled and assessed during the production stages to minimize any possible source-induced
background. If not controlled to sufficiently low levels, the presence of radioactive impurities  in the
\ce{}-\pr{} source could also bias any activity measurement, especially those using global methods such as
calorimetry. Second, source activity and neutrino energy spectrum shape have to be measured to a good accuracy
(at the percent level, see section~\ref{sec:par_source}) in order guarantee a good sensitivity,
especially in the $\Delta m^2 \gtrsim \SI{5}{eV^2}$ region.

\subsection{\label{sec:compo}Composition}
The Russian spent nuclear fuel reprocessing company, Federal State Unitary Enterprise Mayak Production
Association (hereafter \ma{}), has been identified to be the only facility able to deliver a \si{PBq}
scale sealed source of \ce{}. Cerium separation and extraction together with final source packaging and certification
is an operation lasting for several months. The intrinsic abundance of \ce{} along with the extraction process
efficiency requires 5 to \SI{10}{tons} of VVER (the Russian pressurized water reactor technology) spent fuel
to be reprocessed in order to achieve an activity at the PBq level. The first step of the cerium extraction is a
standard reprocessing of nuclear spent fuel components, leading to a lanthanides and minor actinides
concentrate. In a second step, cerium is separated using a complexing displacement chromatography
method~\cite[see][and references therein]{Isolation1,Isolation2}.
No isotopic separation methods are considered in the cerium extraction stage.

The final product will be made of a few kilograms of sintered $\text{CeO}_2$ containing a few tens of grams of
\ce{}~\footnote{\ce{} specific activity is \SI{118}{TBq/g}~\cite{Gerasimov2014}.
Depending on the spent fuel cooling time, the \ce{} mass fraction among all stable isotopes is 
\SIrange{0.1}{1}{\%}.}. The other cerium isotopes,
$^{140,142}\text{Ce}$, are stable. Inevitably, the extraction process leads to small lanthanide and minor
actinide leftovers, that may be penalizing to realize a neutrino experiment. Firstly because these leftovers,
if radioactive, could bias the activity measurement. The level of radioactive impurities must therefore be small
enough (lower than the desired accuracy, i.e. $\lesssim \SI{1}{\%}$) to account for a negligible contribution
to the source activity. This requirement can be achieved using complexing displacement chromatography techniques.
Secondly, because radioactive leftovers can lead to source-induced backgrounds that could degrade the
experimental sensitivity to short baseline oscillations.
A detailed simulation of source-induced backgrounds in a spherical liquid scintillator detector
has been carried out in order to assess the maximum level of such radioactive impurities in the \ce{} source, and is
presented in section~\ref{sec:backgrounds}.

\subsection{Source activity}
As previously stated, a precise knowledge of the source activity is mandatory in order to cover most of the
$\Delta m^2$ region preferred by the short baseline anomalies. As shown in section~\ref{sec:sensitivity}, a
precision at the percent level is necessary and makes any activity measurement very challenging. Direct
source activity measurement, either by $\upbeta$ or $\upgamma$ spectroscopic measurement or isotopic
measurement of the \ce{} concentration is complicated by the thick high-Z shielding used to suppress the
\SI{2.185}{MeV} $\upgamma$ rays released in the \pr{} decay. Using such methods to precisely estimate the
activity then requires source sampling. As such, obtaining a reliable and precise activity measurement further
requires that the $\text{CeO}_2$ source is homogeneous (i.e. the sample is representative of the \ce{}-\pr{}
ANG) and that a measurement of a tiny sample mass with a precision below the percent level can be
achieved. Regarding the previously mentioned measurement methods, $\upbeta$ or $\upgamma$ spectrometer devices
are easier to set up than a mass spectrometer able to handle radioactive materials. Systematic uncertainties
associated to such methods are beyond the scope of this article and shall be further investigated.

Calorimetric measurements offer an attractive alternative to spectroscopic methods. First because the
released heat is directly linked to the source $\upbeta$-decay activity. Second, because it allows the
measurement of the source activity without the need for sampling. In such a case, the design of an
instrumented calorimeter, dedicated to fitting the source and its associated shielding, is mandatory. Beyond
the heat power measurement, great care must be taken in the calculation of the power-to-activity conversion
constant. This quantity is calculated using the available information on the different \ce{} and \pr{}
$\upbeta$ branches from nuclear databases, such as branching ratios and $\upbeta$ mean energies. The
uncertainty on those quantities could possibly limit the final activity measurement precision. Using
information taken from the Chart of Nuclides~\cite{NNDC}, the computation of the power-to-activity conversion
constant for the \ce{}-\pr{} couple gives \SI{216.0(12)}{W/PBq} (\SI{0.56}{\%} uncertainty). The \SI{0.56}{\%}
uncertainty is here dominated by the uncertainty on the branching ratio of the main \pr{} $\upbeta$ branch.
Furthermore, the $\upbeta^-$ model used to compute the mean energy per branch is the LOGFT code~\cite{LOGFT},
which treats any $\upbeta$ transitions as allowed transitions~\cite{LOGFT-readme}, whereas the main $\upbeta$
branches of the \ce{}-\pr{} couple are non-unique first forbidden transitions. As explained in the next
section, the \ce{}-\pr{} energy spectrum shape modeling and $\upbeta$ spectroscopic
measurement will help improving the precision on this conversion constant. The current $\upbeta$ spectrum
shape modeling (see next section) gives \SI{215.6(13)}{W/PBq} (\SI{0.60}{\%} uncertainty) for the
power-to-activity conversion constant. The branching ratio uncertainties are still dominant with a small
additional source of uncertainties coming from the corrections applied to Fermi's theory.

\subsection{\label{sec:betaspectrummodeling}Spectral shape: modeling and $\upbeta$ spectrometry}
\begin{figure*}[ht]
\centering \includegraphics[width=.49\linewidth]{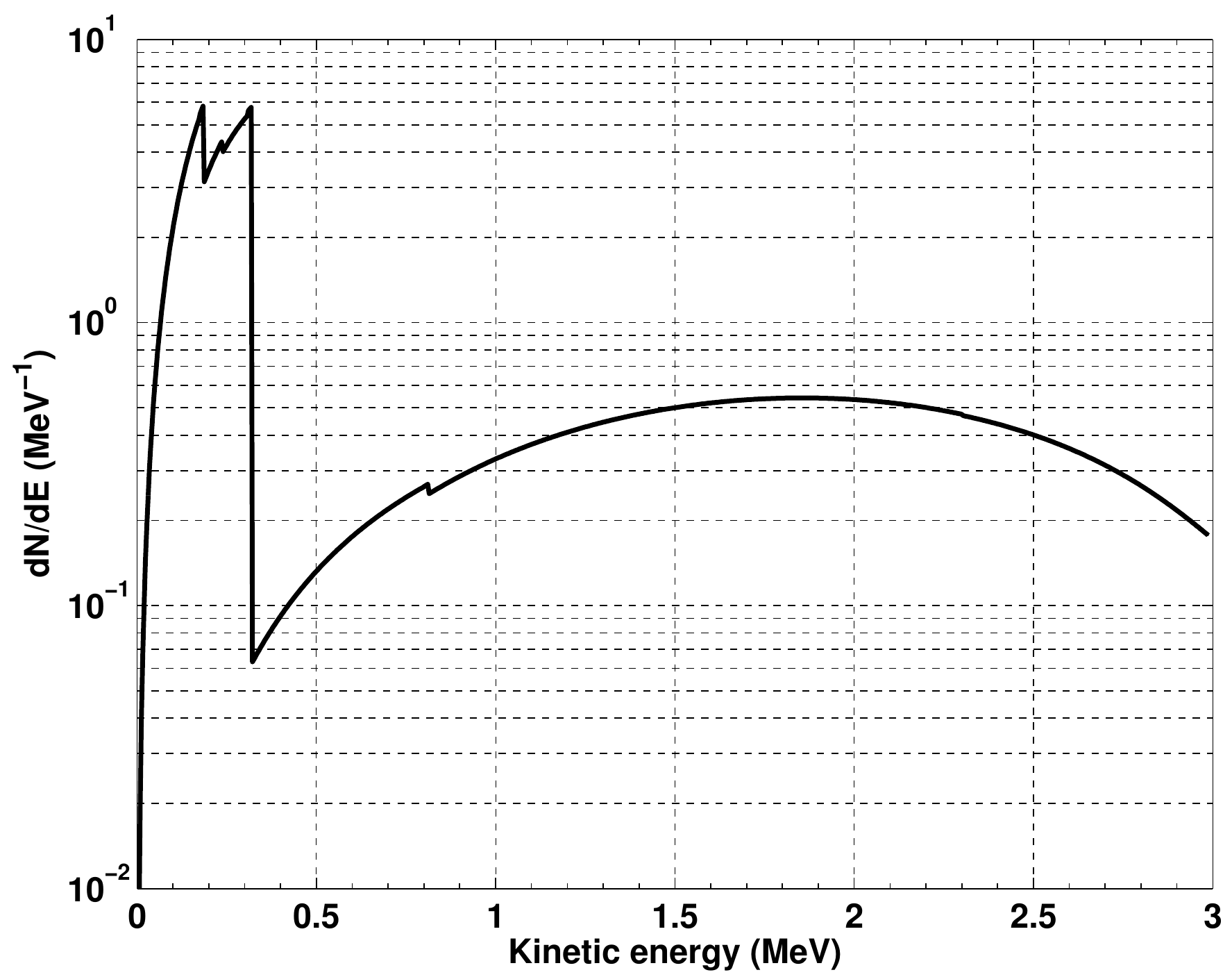}
\centering \includegraphics[width=.49\linewidth]{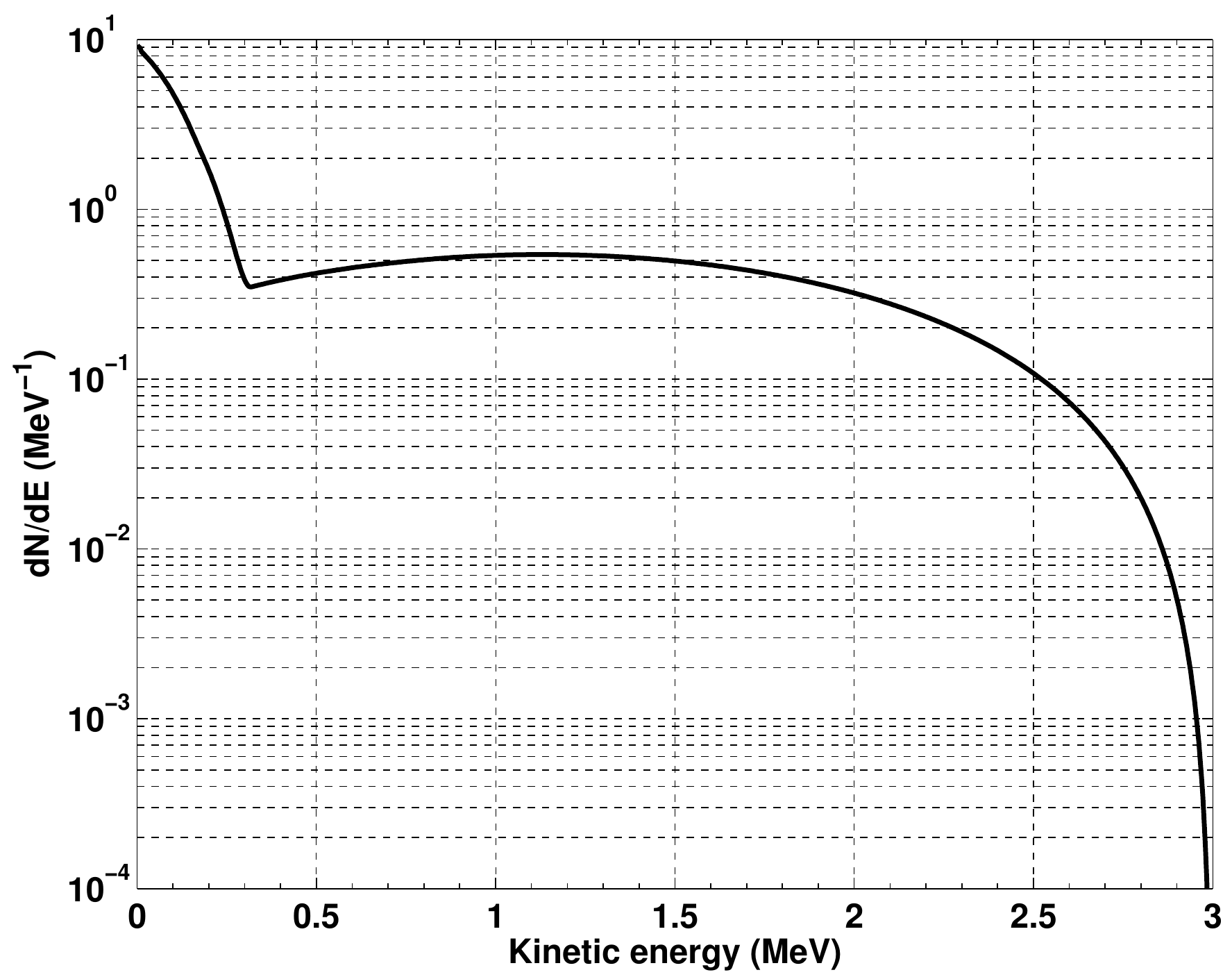}
\caption{\label{fig:Combined_nu_spectrum} Estimation of \anu{} spectrum without neutrino oscillations (left)
and of $\upbeta$ spectrum (right) of the \ce{}-\pr{} couple. Both spectra are normalized to two decays.}
\end{figure*}

Both the \ce{}-\pr{} source $\upbeta^-$ and \anu{} energy spectra have to be accurately known.
As explained in the previous section, calorimetric methods require a very good accuracy on the
power-to-activity conversion factor to reach a percent level precision on the source activity. This conversion
factor is related to the mean energy per decay released by the $\upbeta$ particles and is sensitive to the
spectrum shape. A precise knowledge of the full \ce{}-\pr{} $\upbeta$ spectrum is then necessary to compute
this conversion factor to a good precision, which is critical for estimating the ANG activity and then
predicting the number of IBD interactions in the detector. Moreover, the number of expected events depends on
the fraction of \anu{} emitted above the IBD threshold. Therefore, the accuracy achieved by a rate+shape
analysis (see section ~\ref{sec:sens}) will strongly depend on the \anu{} spectrum shape uncertainty.
To a lesser extent, any uncertainties on the \anu{} energy spectrum shape will also degrade the experiment
free rate sensitivity (see section ~\ref{sec:sens}) to any distortions caused by neutrino oscillations at
short baselines. However this effect is reduced by the highly characteristic oscillating pattern of the
expected signal.

\begin{figure}[ht]
\centering \includegraphics[width=\linewidth]{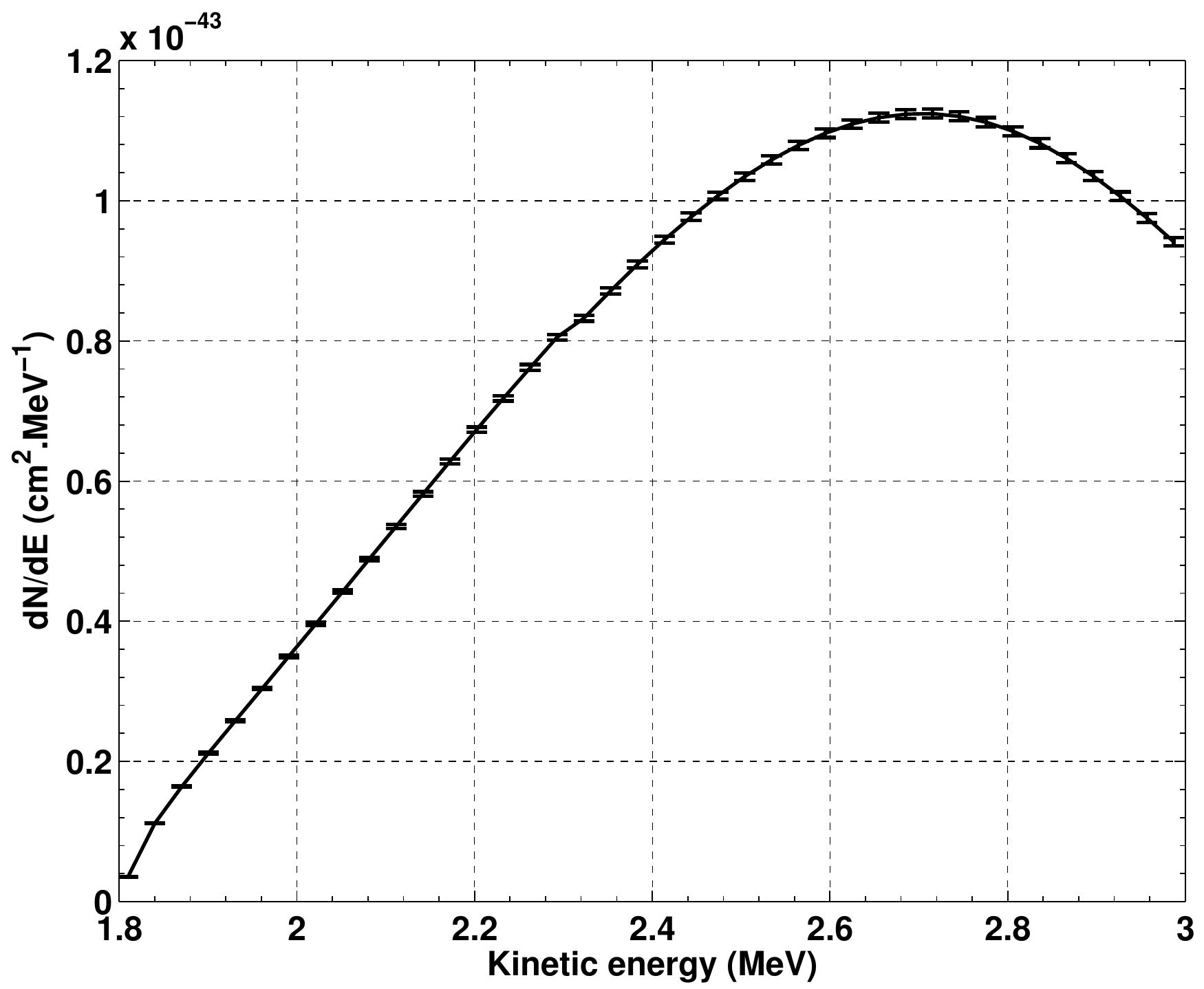}
\caption{\label{fig:pr_detectedflux} Differential observable neutrino spectrum from \pr{}
folded with the IBD cross-section.}
\end{figure}

The \ce{}-\pr{} $\upbeta$ and \anu{} spectra are a combination of several $\upbeta$ branches. To a very good
approximation, they are linked at the $\upbeta$ branch level through the following energy conservation
relationship $E_{\upbeta^-} = Q_{\upbeta} - E_{\anu{}}$, where $E_{\upbeta^-}$, $Q_{\upbeta}$ and $E_{\anu{}}$
are the $\upbeta$ particle kinetic energy, the $\upbeta$ branch endpoint energy and the \anu{} energy, respectively.
In addition to the $\upbeta$-transitions presented on figure~\ref{fig:144Ce-144Pr}, \pr{} presents seven
secondary branches with branching ratios smaller than \SI{0.01}{\%}~\cite{NNDC}. Among all possible \ce{} and
\pr{} $\upbeta$ transitions, only two transitions exhibit endpoint energies larger than the IBD reaction
energy threshold. They come from the decay of \pr{} and total \SI{98.94}{\%}  its decays. Although the
other \ce{} and \pr{} $\upbeta$ branches are irrelevant regarding \anu{} detection, they have to be carefully
studied and measured for the calculation of the power-to-activity conversion factor.

A calculation of the full \anu{} spectrum, assuming that the \ce{}-\pr{} couple is at secular equilibrium, is
shown on figure \ref{fig:Combined_nu_spectrum}. The detected spectrum is shown on
figure~\ref{fig:pr_detectedflux}. This calculation has been done using Fermi's theory of $\upbeta$
decay~\cite{Mueller2011,Huber2011} of an infinitely massive point-like nucleus, corrected for various effects
such as finite size and mass of the nucleus (considering both electromagnetic and weak interaction), nucleus
recoil effect on the phase space and Coulomb field~\cite{Wilkinson1990}, first order radiative
corrections~\cite{Sirlin2011}, screening effect of the atomic electrons~\cite{Behrens1982} and weak magnetism
effect~\cite{Huber2011,Huber2012}.

Each effect represents up to a few percents correction to the \anu{} spectrum shape, excepted the nucleus
recoil effect which has a negligible impact for the \ce{} and \pr{} heavy nuclei. These corrections are
generally known with a sub-percent precision, except the weak magnetism effect, for which the few available
calculations are uncertain and available data are too sparse in order to perform a reliable comparison.
However, the main \pr{} $\upbeta$ branch (\SI{97.9}{\%} branching ratio) is not affected by weak magnetism as
it is a $0^{-} \rightarrow 0^{+}$ non-unique first forbidden transition with no angular momentum
change~\cite{Calaprice2014,Hayes2014}. The \anu{} spectrum final uncertainty budget is then dominated by the
screening correction, which dominates in the high energy part of the spectrum, and the shape factors of the
non-unique first forbidden $\upbeta$ branches occurring in the \ce{} and \pr{} decay schemes (see figure
~\ref{fig:144Ce-144Pr}). The final uncertainty budget on the \anu{} spectrum modeling is of the order of
\SI{1}{\%}.
The impact on the number of expected events has been checked by varying the spectrum shape according to this
uncertainty budget. Such a distortion of the spectrum would lead to a \SI{0.5}{\%} variation in the events
number and to a \SI{0.2}{\%}  variation in the power-to-activity conversion factor.

As concerns the modeling of the \ce{}-\pr{} $\upbeta$ spectrum, the non-unique first forbidden $\upbeta$
branch shape factors and corrective terms to Fermi's theory have the same amplitudes and uncertainties than
those used to model the \anu{} spectrum, except the radiative corrections. Radiative corrections are not
symmetric between the $\upbeta$ and \anu{} spectra. Radiative corrections to the electron spectrum modeling
show larger uncertainties and are more difficult to estimate than those corresponding to the \anu{} spectrum
modeling. The radiative correction to the electron spectrum modeling can rise up to a few
percents~\cite{Behrens1982}.

A precise measurement of the \ce{}-\pr{} $\upbeta$ spectrum is then necessary to further reduce the shape
uncertainties. The $\upbeta$ spectrum modeling presented previously will be used to interpret the data, and
possibly to constrain the most uncertain corrections to the first-order Fermi's theory. A few challenges arise
though for a precise $\upbeta$ spectroscopic measurement of the \ce{}-\pr{} couple. For example, the short
period of \pr{} makes it difficult to measure its associated $\upbeta$ spectrum independently of \ce{} 
$\upbeta$ spectrum, especially at low energies (i.e. at energies greater than the IBD reaction threshold in 
the \anu{} case) where the two spectra overlap. A chemical separation of \pr{} from \ce{} can be performed to 
circumvent this issue. Because of the short \pr{} half-life it requires a dedicated separating setup, such as
chromatographic columns, installed in the vicinity of the $\upbeta$ spectrometer.

Several methods are considered to perform the \ce{}-\pr{} $\upbeta$ spectrometry. They are complementary
in the sense that they are sensitive to different instrumental effects. The most precise
measurement would be achieved using a magnetic spectrometer. However, such a device could be difficult to
find. Semiconductor (silicon or germanium) or plastic scintillator detectors could also be used, provided they are
thick enough to avoid the most energetic \SI{3}{MeV} electrons to escape. The interpretation of these data could 
be  challenging because of backscattering effects at low energies~\cite{KNOLL}. Finally,
liquid scintillator devices can also be used to measure radionuclide $\upbeta$ spectra. Using such an apparatus
would allow the \ce{}-\pr{} mixture to be added into the detection volume, preventing backscattering effects.
However, energy resolution would be limited by the scintillator light yield and the light collection
efficiency. Moreover, the device energy response would have to be precisely modeled, especially to correct
for the escape of the most energetic $\upbeta$ particles out of the scintillator volume.

\section{\label{sec:neutrino_detectors}Neutrino detectors and deployment sites}
Any neutrino detector suitable for the search of meter-scaled oscillations should meet the following
requirements: kiloton-scale detection volume, target liquid rich in free protons, low background event rate
in the MeV-energy range (consequently located deep underground), able to lower the energy threshold below
\SI{1}{MeV} (minimum visible energy of an IBD event).

The kiloton scale detection volume is necessary to compensate the relatively low \anu{} flux of an ANG with a
few~\si{PBq} activity. The low energy threshold requirement prevents the use of current water Cerenkov
detectors~\footnote{The water Cerenkov light yield is about~\SI{300}{photons/MeV}, compared to roughly
\SI{1e4}{photons/MeV} for typical liquid scintillator.}, leaving liquid scintillator based detectors the only
suitable candidates. As mentioned in section~\ref{sec:exp_concept}, these detectors are namely the KamLAND
detector, located in the Japanese Kamioka Mine~\cite{Eguchi:2002dm}, the Borexino detector, located in the
Italian Gran Sasso National Laboratory (LNGS)~\cite{Alimonti:2008gc} and the soon to be commissioned SNO+
detector, in the Canadian Sudbury mine laboratory (SNOLAB)~\cite{Chen:2008un}.

\begin{figure*}[ht]
\centering 
\includegraphics[width=0.49\linewidth]{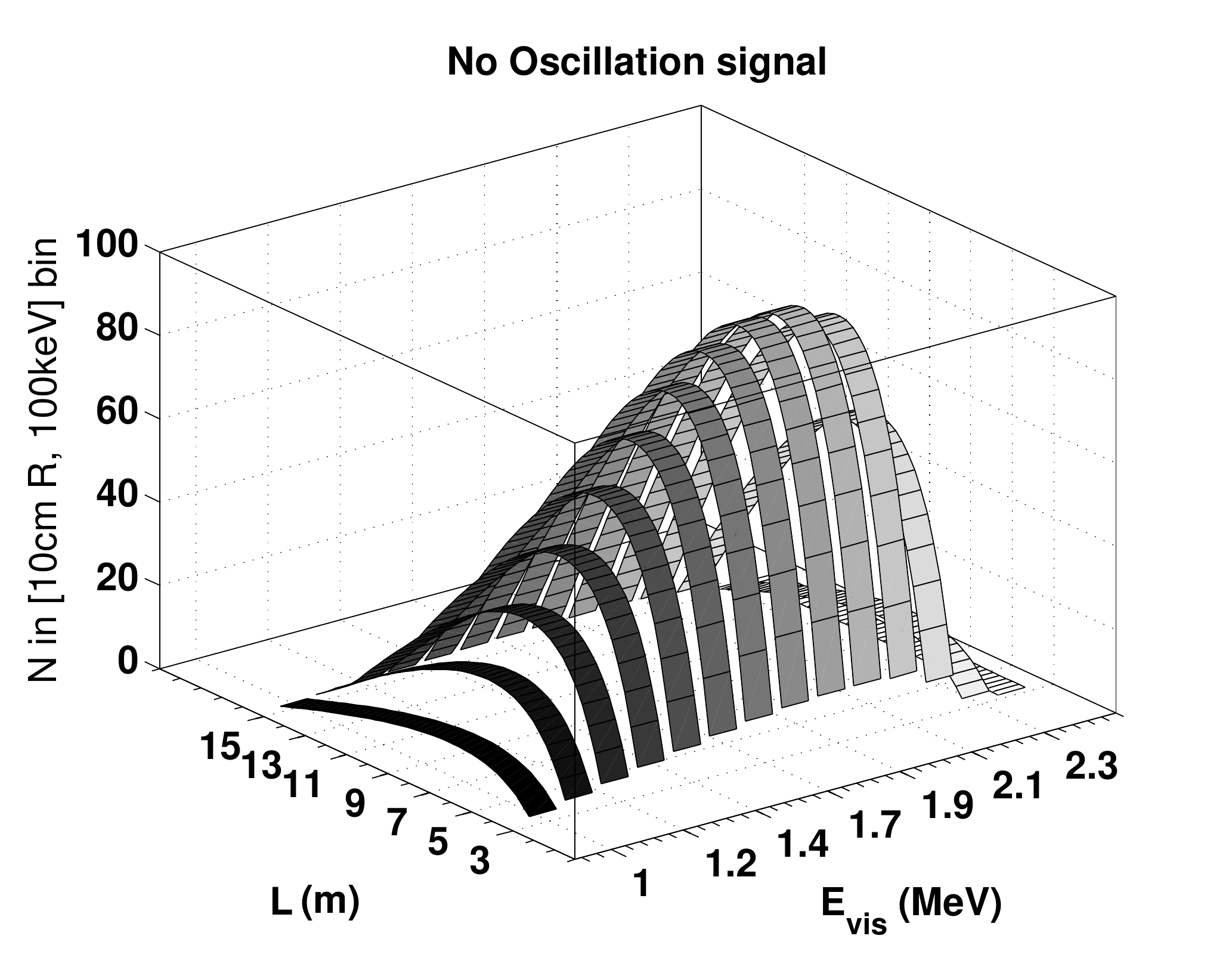}
\includegraphics[width=0.49\linewidth]{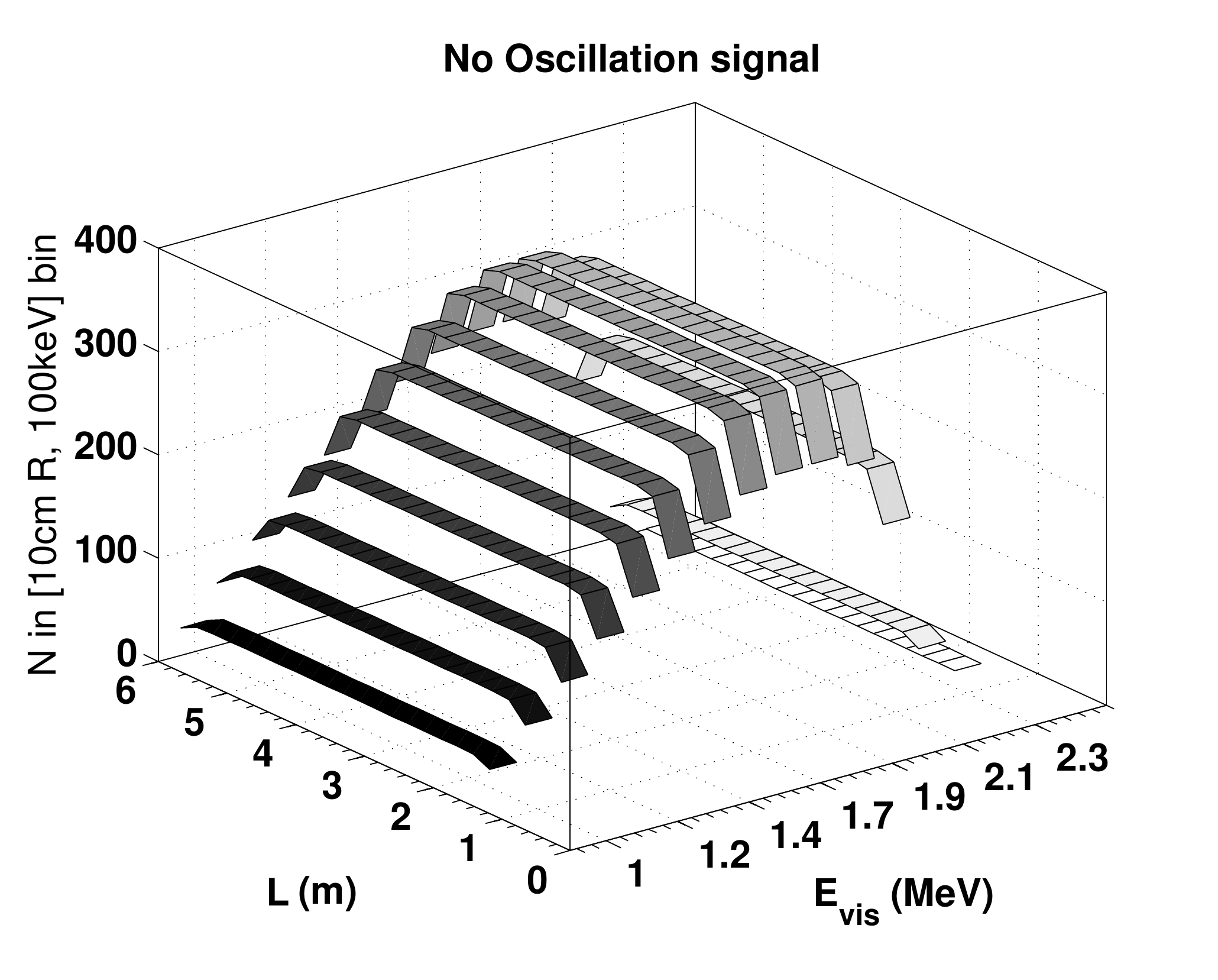}
\caption{\label{fig:Bisignal} Expected signal as a function of both visible energy $\mathrm{E_{\rm vis}}$ and
baseline L, in the no-oscillation hypothesis, for a ANG positioned either outside (left) or at the center
(right) of a liquid scintillator spherical detector. See text for further details.}
\end{figure*}

These detectors share a similar layout, based on nested enclosures. Starting from the outer part, the first
enclosure is a large cylindrical water tank holding PMTs. It is used as a muon veto and as a first
shielding against natural radioactivity and fast neutrons induced by cosmic muons. Going inward, a second
spherical vessel is nested within the water tank and is also mounted with a large number of PMTs. The second vessel
itself includes a third transparent acrylic vessel (SNO+), or nylon vessel (Borexino, KamLAND), which defines
the target volume for the detection of neutrinos. It is filled up with hydrogen-rich liquid scintillator. The
region between the second and third vessels is filled up with a non-scintillating and transparent material,
and is called the buffer region. The buffer region separates the PMTs from the target, ensuring a good optical
coupling between the scintillator and PMTs. This design enhances the uniformity of the detector response, and
further shields the target volume against the radioactivity of the surrounding materials.

The large PMT coverage of the KamLAND, Borexino and SNO+ detectors, associated with their good liquid
scintillator properties, allow to achieve an energy resolution of about \Eres{} at \SI{1}{MeV} and a vertex
resolution better than \Lres{}, making them very interesting detectors for performing a short baseline
neutrino experiment. The main differences come from their target mass, and also the
possibility and technical feasibility of deploying a radioactive source in their vicinities.
This last point is closely examined in the next paragraphs.

Considering a deployment scenario at KamLAND, the \ce{} ANG could be positioned in the water tank close to the
inner detector volume, \SI{10}{m} away from the target center. A second solution, technically easier and
faster than the previous one would consist in placing the source in an existing storage room, \SI{12}{m} away
from the target center. Such a deployment scenario, called CeLAND, has been investigated in details, and is
described in reference~\cite{CeLAND_proposal}. In a second phase, the ANG could be relocated at the center of
the target if hints for a signal show up in the first months of data taking. However, such a deployment
scenario would require further technical challenges to be solved. For example, it would require increasing the
high-Z shielding thickness ($\sim$ 30-35 cm) with respect to the first phase (see for example the preliminary
study presented in~\cite{Cribier:2011fv}). It would also need to solve additional technical safety and
deployment issues, and more importantly to guarantee a level of gamma and neutron emitting contaminants in the
\ce{}-\pr{} source much lower than those previously discussed.

A deployment of a \ce{}-\pr{} ANG next to the main stainless steel vessel of the Borexino detector is actually
planned at the end of 2015. Such an experiment is called CeSOX~\cite{SOX}. The ANG would be located in the pit
underneath the detector, at a distance of \SI{8.25}{m} from the target center. Two possible experimental
configurations are worth considering in the deployment of an ANG near Borexino. The first experimental
configuration uses the Borexino detector as it is today, i.e. with a fiducial volume defined by a radius of
\SI{4.25}{m} around the detector center. The second experimental configuration assumes an enlarged detection
volume, achieved by doping the buffer oil surrounding the primary target with fluors and wavelength shifters.
The detection volume radius would reach \SI{5.5}{m}, increasing the target mass by a factor \num{2.2}.
Finally, the deployment of the \ce{}-\pr{} source at the center of the Borexino detector is presently
disfavoured because of space and mechanical constraints at the detector chimney level. Nevertheless, such a
deployment could be realized after the refurbishment of the detector neck in a long term.

A third option concerns the forthcoming SNO+ detector that will start commissioning in the next months. For
our sensitivity study, a deployment of the ANG close to the inner detector stainless steel tank, \SI{10}{m}
away from the target center, will be considered. Among the three detectors being considered SNO+ appears to be
the most suitable for a deployment within the target scintillating volume, because of its wide chimney of
\SI{1.2}{m} diameter.

The three detector main characteristics are detailed in Table~\ref{tab:detector}. KamLAND offers the highest
density of free protons, which is \SI{25}{\%} higher than in Borexino, and \SI{6}{\%} higher than in SNO+. If
no fiducial volume cut is applied (for example to suppress source-induced backgrounds), the detection volume
available in KamLAND is a factor 4.5 larger than in Borexino, and 1.35 larger than in SNO+. In the case of
Borexino, this volume difference could be partially compensated by deploying the \ce{} ANG closer to the
detector center. As an example, deploying a \SI{3.7}{PBq} \ce{} ANG at the locations mentioned above for each
detector, and for  \SI{1.5}{years} of data taking, leads to \num{2.9e4}, \num{9.2e3} and
\num{2.1e4} IBD interactions in KamLAND, Borexino (R=\SI{4.25}{m}), and SNO+, respectively. Increasing the
Borexino detection volume such that R=\SI{5.5}{m} in would give \num{2.1e4} IBD events.

\begin{table}[!ht] {
\newcounter{A} \setcounter{A}{1}
\newcounter{B} \setcounter{B}{2}
\begin{center}
 \begin{tabular}{>{\centering}m{2cm} >{\centering}m{1cm}
>{\centering}m{1cm}>{\centering}m{1.8cm} >{\centering}m{1.5cm} }
  \toprule
  Detector & mass (\si{ton}) & radius (\si{m}) & density (\SI{e28}{H/m^3})\protect
\footnotemark[\theA]& \# of H (\num{e31})\protect \footnotemark[\theA] \tabularnewline
\colrule
  \rule{0pt}{2.5ex}
  KamLAND  & 1000 & 6.5 & 6.60 & 7.6\tabularnewline
  SNO+     & 780  & 6.0 & 6.24 & 5.6\tabularnewline
  Borexino geo-$\nu$ & 280 \protect \footnotemark[\theB] & 4.25 & 5.3 & 1.7
\tabularnewline
  Borexino extended & 415 \protect \footnotemark[\theB] & 5.5 & 5.3 & 3.7\tabularnewline
  Generic & 600  & 5.5 & 6.24 & 4.3\tabularnewline  \botrule
 \end{tabular}
\footnotetext[\theA]{H stands for hydrogen nuclei.}
\footnotetext[\theB]{Data extrapolated from \SI{100}{ton} in \SI{3}{m} radius.}
\caption{\label{tab:detector} The main characteristics of the liquid scintillator detectors suitable to look
for short baseline neutrino oscillations.}
\end{center}
}
\end{table}

Finally, it is worth mentioning that the detector background conditions will not be identical because of
different overburdens and contamination by structure material radioactive impurities. However, the detector
induced-backgrounds are expected to be very low compared to the \ce{} ANG signal. If necessary, they can be
well identified and measured before or after the ANG deployment. Thus in the following studies,
detector-induced backgrounds will be neglected.

\section{\label{sec:sensitivity}Experimental sensitivity to short baseline oscillations}
This section details the expected neutrino signal modeling and sensitivity to short baseline oscillations.
The impact of several source and detector related experimental parameters on the sensitivity is also studied.

\begin{figure}[ht]
\centering \includegraphics[width=\linewidth]{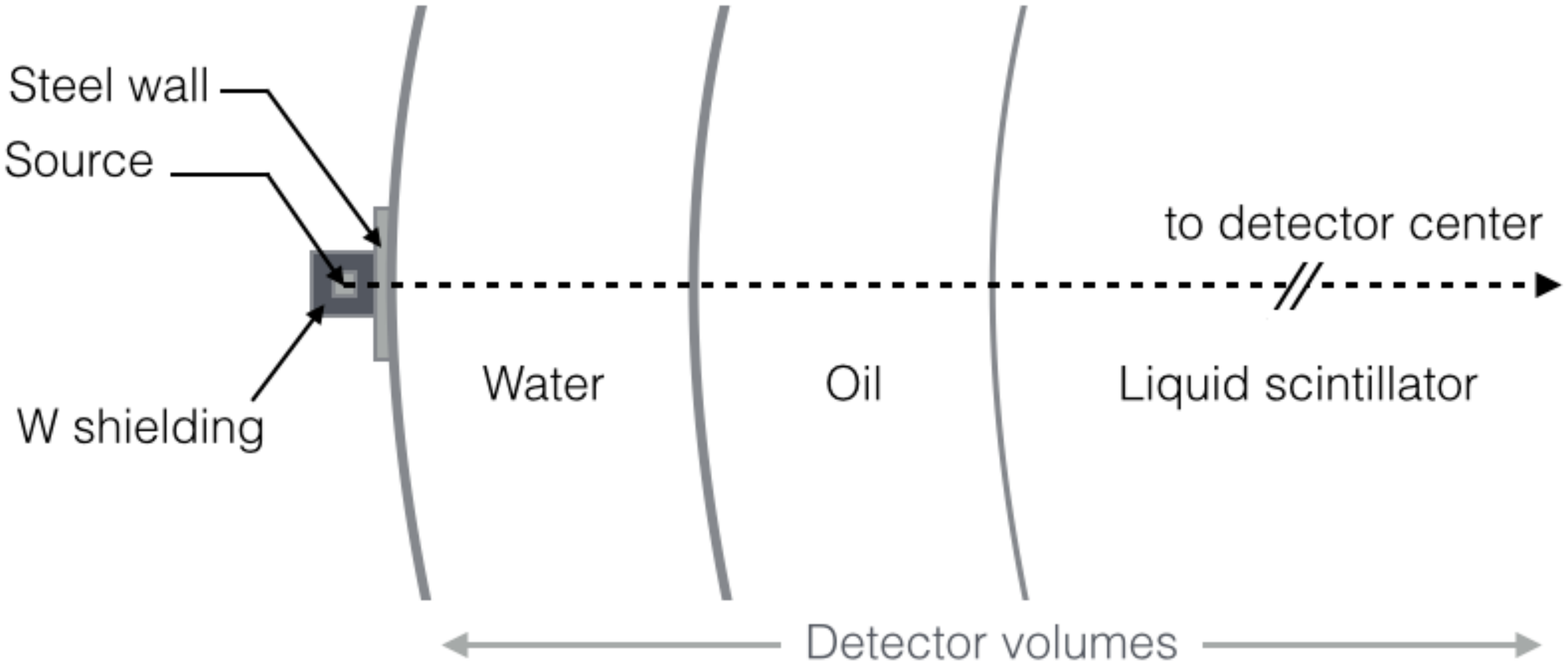}
\caption{\label{fig:genericsetup} Sketch of the generic experimental setup.}
\end{figure}

Unless otherwise stated, the study assumes a generic experimental configuration, where a \SI{3.7}{PBq}
\ce{}-\pr{} ANG is deployed \SI{10}{m} away from the center of a spherical liquid scintillator detector. The
detector is made of an active target of \SI{5.5}{m} radius, with a buffer region and muon veto which are both
\SI{2}{m} thick. The ANG is positioned under the detector below a \SI{10}{cm} thick steel plate. It is a
cylindrical capsule of \SI{14}{cm} radius and height, filled up with $\text{CeO}_2$, and inserted into a
\SI{19}{cm} thick tungsten shielding. A sketch of this experimental setup is shown in figure 
\ref{fig:genericsetup}.

The target liquid scintillator is assumed to be Linear AlkylBenzene (generally known as LAB),
whose density is \SI{0.86}{g/cm^3}, leading to a target mass of \SI{600}{t}.
The detector has a $\SI{5}{\%}/\sqrt{E}$ energy resolution and a \SI{15}{cm} vertex resolution.
No fiducial volume cut is assumed in the signal and sensitivity predictions. In the no-oscillation scenario
\num{1.6e4} IBD interactions would be expected with data taking time of \SI{1.5}{years}. Expected ANG
induced backgrounds are discussed in section \ref{sec:backgrounds} and summarized in table
\ref{tab:signal_bkg}.

\subsection{\label{sec:signal}Expected signal}
The expected number of \anu{} events $N_{\anu{}}$ in a volume element $\ud^3 \mathscr{V}_\mathrm{det}$ located
at a distance L from a point-like ANG of initial activity $\mathcal{A}_0$, during a time $\ud t$ and in an
energy interval $\ud E$, can be expressed as follows:
\begin{multline}
\label{eq:nu_rate_ext}
\dfrac{\ud^5 N_{\anu{}}}{\ud t \, \ud E \, \ud^3 \mathscr{V}_\mathrm{det}} = \mathcal{A}_0 \s e^{-t \s
\lambda_{\text{Ce}}} \s \eta_p \s \varepsilon \s \dfrac{1}{4 \pi L^2} \s \sigma_\mathrm{IBD}(E) \s
S_{\text{Ce}}(E)\\
\times \s \mathcal{P}(L,E),
\end{multline}
where $\lambda_{\text{Ce}}$ is the \ce{} decay constant (s$^{-1}$), $\eta_p$ is the free proton density and
$\varepsilon$ is the detection efficiency, which can be position-dependent. $\sigma_\mathrm{IBD}$ stands for
the IBD reaction cross-section, and $S_{\text{Ce}}$ is the \ce{}-\pr{} \anu{} spectrum as computed with the
modeling discussed in section~\ref{sec:betaspectrummodeling}. The interpretation of the short baseline
anomalies in terms of neutrino oscillation prefers new mass splittings $\Dmn \sim \SI{1}{eV^2} \gg \Delta
m_\mathrm{sol}^2,\,\Delta m_\mathrm{atm}^2$. The $\mathcal{P}(L,E)$ survival probability of \anu{} can then be
written assuming a 2-flavor oscillation scenario:
\begin{equation}
\label{eq:survival_proba}
\mathcal{P}(L,E) = 1-\stn \s \sin^2 \! \left( \dfrac{c^3}{\hbar}
\dfrac{\Dmn \, L}{2 E} \right),
\end{equation}
where \tn{} is the new mixing angle associated to the \Dmn{} new mass splitting.

If the ANG is deployed at the center of the detector, equation~\ref{eq:nu_rate_ext} reduces to:
\begin{equation}
\label{eq:nu_rate_int}
\dfrac{\ud^3 N_{\anu{}}}{\ud t \, \ud E \, \ud r} = \mathcal{A}_0 \s e^{-t \s \lambda_{\text{Ce}}} \s \eta_p \s
\varepsilon \s \sigma_\mathrm{IBD}(E) \s S_{\text{Ce}}(E) \s \mathcal{P}(r,E)
\end{equation}
where the integral over the zenithal and azimuthal coordinates of the detector volume element
$\ud^3\mathscr{V}_\mathrm{det}$ (expressed in polar coordinates) has been performed.
Equation~\ref{eq:nu_rate_int} translates the fact that in the no-oscillation hypothesis, the count rate would
be constant as a function of the  distance to the center of the detector~$r$.

\begin{figure}[h]
\centering
\includegraphics[width=\linewidth]{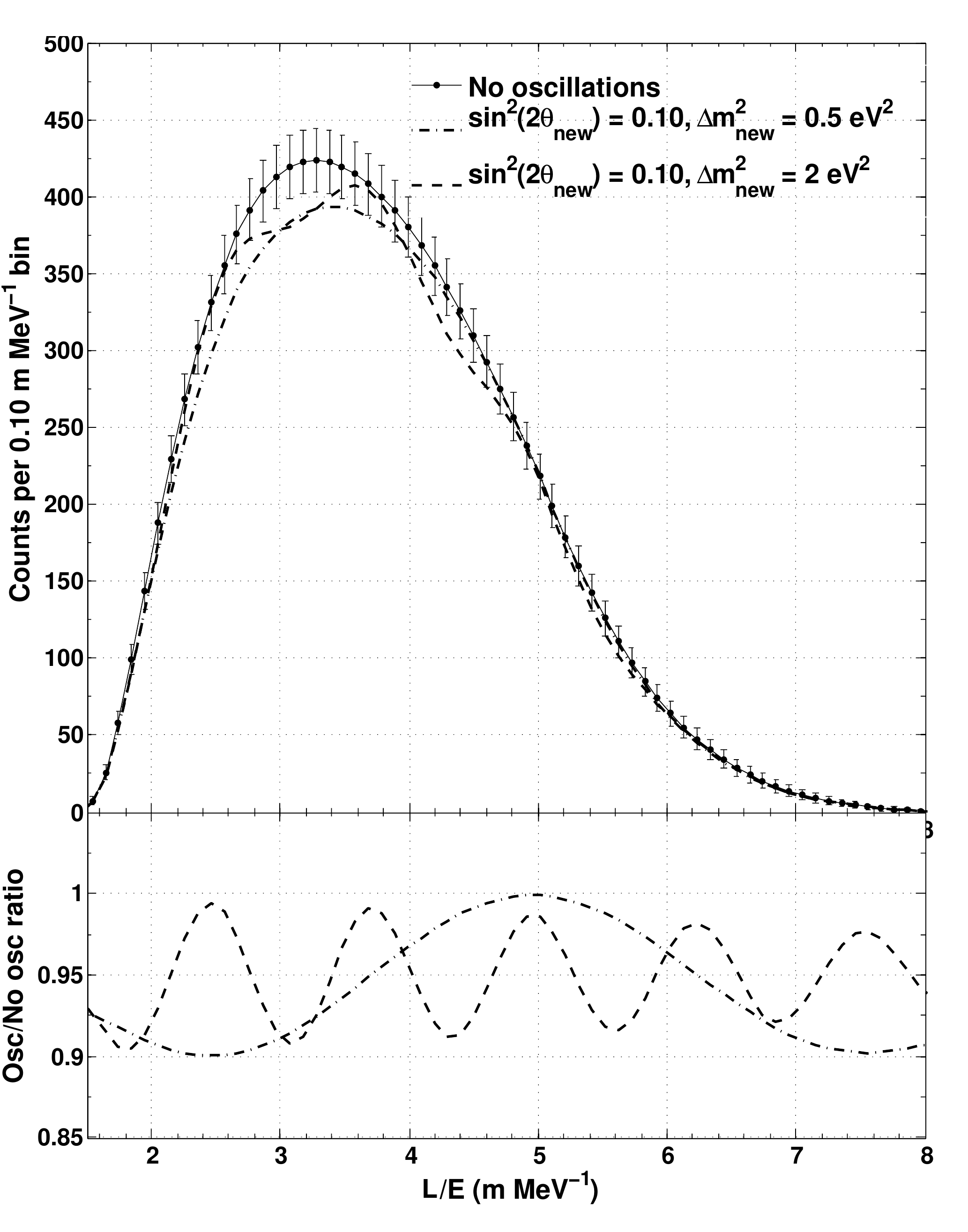}
\caption{\label{fig:SignalLoE} Upper panel shows the expected signal as a function of $\mathrm{L/E}$. Error
bars includes statistical uncertainties corresponding to \SI{1.5}{years} of data taking and a \SI{1.5}{\%}
uncertainty on the source activity. The no-oscillation scenario is compared to two oscillation scenarios, with
squared mass splittings $\Dmn=0.5$ and \SI{2.0}{eV^2} and a $\stn=0.1$ mixing parameter. Lower panel shows the 
ratio of the oscillated spectra to the non-oscillated spectrum.}
\end{figure}

\begin{figure*}[ht!]
\centering
\includegraphics[width=0.49\linewidth]{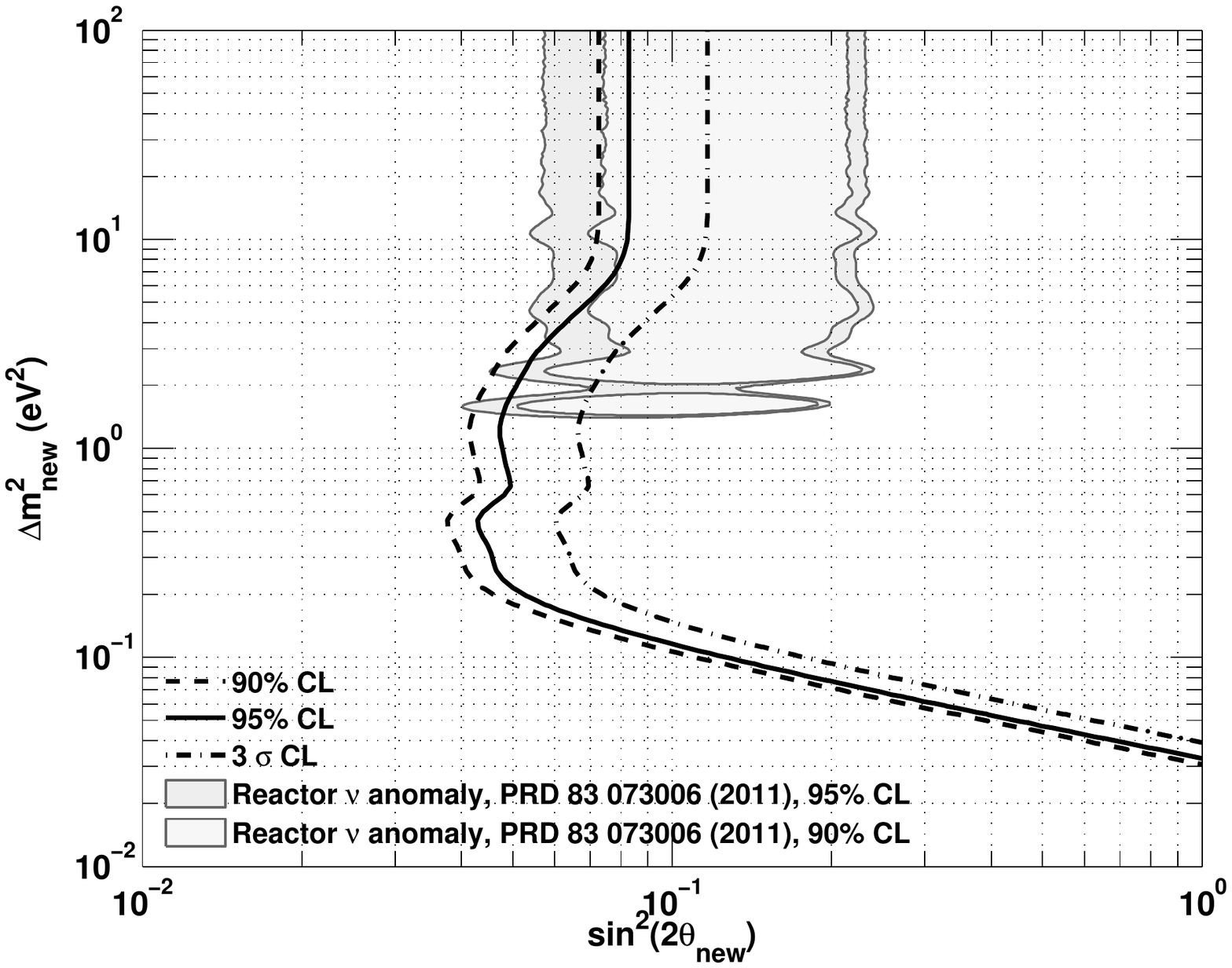}
 \includegraphics[width=0.49\linewidth]{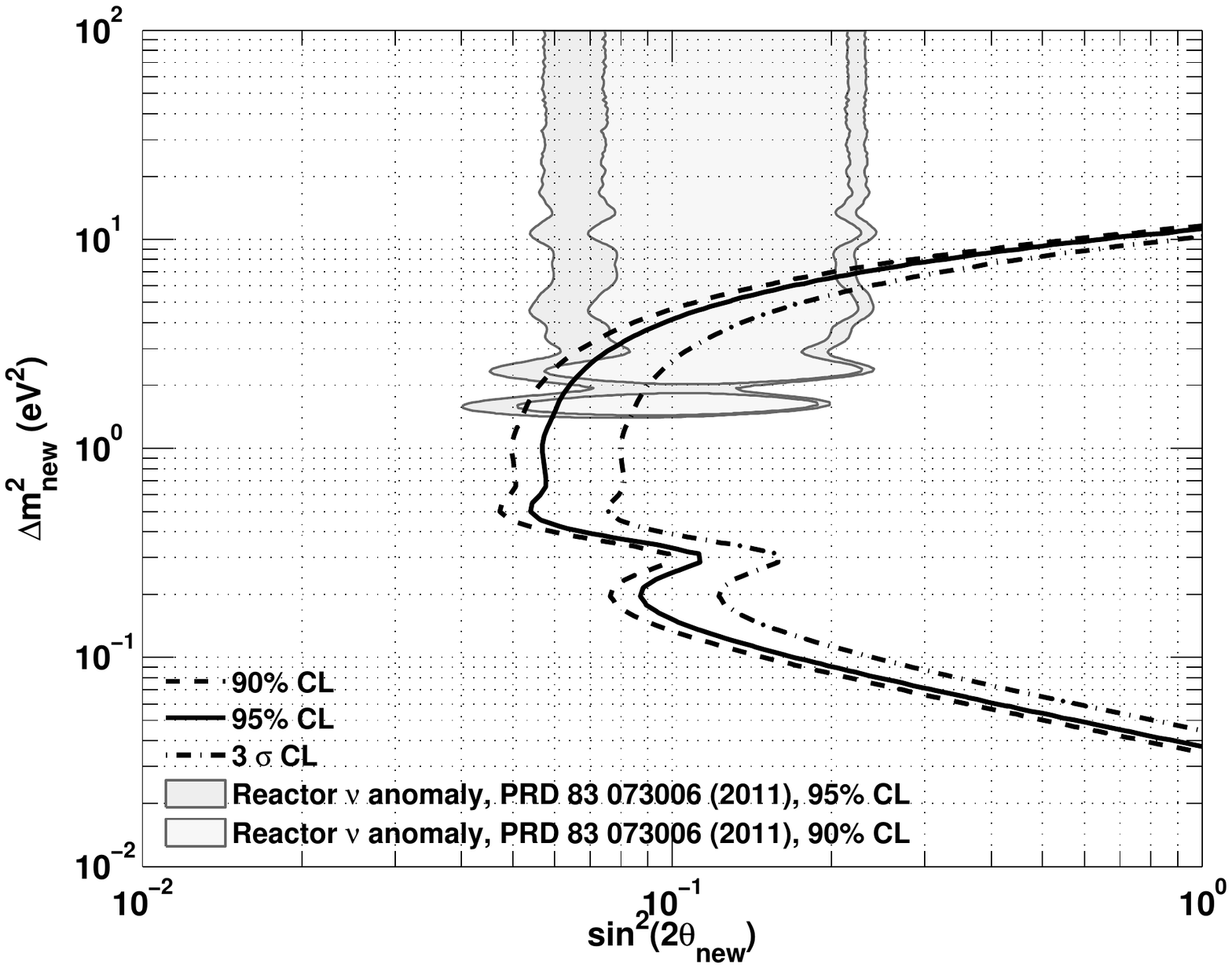}
\caption{\label{fig:ref}
Sensitivity of our generic reference experiment at \SI{90}{\%}, \SI{95}{\%}, and \SI{99.73}{\%} C.L.
The inclusion contours of the reactor antineutrino anomaly are shown in the grey shaded areas.
Left: Rate+shape sensitivity.
Right: Free rate sensitivity.
\\Unless explicitly mentioned, default parameters for sensitivity contours include a \ce{}-\pr{} source of 
\SI{3.7}{PBq} deployed at \SI{10}{m} from the detector center for \SI{1.5}{year}, with \SI{1.5}{\%} 
uncertainty on normalization, vertex and energy resolution at \SI{1}{MeV} of \SI{15}{\%} and \SI{5}{\%} 
respectively, and contours given at the \SI{95}{\%} C.L.
}
\end{figure*}

The detector vertex and energy resolution must be taken into account when computing binned spectra. The
computation of such spectra is done by integrating equation~\ref{eq:nu_rate_ext} or
equation~\ref{eq:nu_rate_int}, smeared with Gaussian vertex and energy resolution functions. The expected
signal in the absence of oscillations, binned both as a function of L and E, is shown on
figure~\ref{fig:Bisignal} for the generic experimental configuration described above. The L and E binned
spectrum corresponding to a source positioned at the detector center is also shown for comparison. Because the
survival oscillation probability depends only on the L/E ratio, the expected signal as a function of L/E has
also been calculated and is shown on figure~\ref{fig:SignalLoE}, for different oscillation scenarii, along
with the corresponding ``oscillated over non-oscillated" ratios. Computing such a ratio would lead to the
\anu{} survival probability (equation~\ref{eq:survival_proba}) for an ideal detector with no vertex and energy
resolutions. The lower panel of figure~\ref{fig:SignalLoE} shows that the corresponding oscillations are
damped with respect to the raw survival probability, because of the detector vertex and energy resolutions.
Damping of the oscillation pattern is especially visible at high \Dmn{} where the oscillation wiggles grow
smaller with higher $L/E$.

\subsection{\label{sec:sens}Sensitivity and discovery potential}
The sensitivity to short baseline oscillations is evaluated by comparing the observed event rate, binned as a
function of both energy and distance, with respect to the expected distribution in the presence of
oscillations. The following $\chi^2$ function is used to test the hypothesis of no oscillation and for
(\Dmn{}, \tn{}) parameter estimation:
\begin{equation}
\chi^2 = \sum_{i,j} \dfrac{ \left( N_{i,j}^\mathrm{obs} - (1+\alpha)
N_{i,j}^\mathrm{exp} \right)^2}
{ N_{i,j}^\mathrm{exp}
 }
 + \left(\dfrac{\alpha}{\mathrm{\sigma_N}}\right)^2,
\end{equation}
where $N_{i,j}^\mathrm{obs}$ and $N_{i,j}^\mathrm{exp}$ are the observed and expected number of IBD events
in the $i^\mathrm{th}$ energy bin and $j^\mathrm{th}$ distance bin, respectively. As explained in the
previous section, they are computed according to equation~\ref{eq:nu_rate_ext} taking into account the
detector energy and vertex resolution. The nuisance parameter $\mathrm{\alpha}$ allows the signal
normalization to vary within its associated uncertainty $\mathrm{\sigma_N}$. The normalization uncertainty is
here assumed to come from the uncertainty on the source initial activity ${\cal{A}}_0$. Setting the source
activity uncertainty $\mathrm{\sigma_N}$ to $\infty$ allows an overall floating normalization and a sensitivity
study which mostly uses shape distortions to look for oscillations. This is the so-called ``free rate'' analysis.
Setting $\mathrm{\sigma_N}$ to the precision achieved by any activity measurement performed prior to the final
experiment allows to use a rate information in addition to the information brought by a spectrum shape
deformation. This is the so-called ``rate+shape'' analysis.

\begin{figure*}[ht]
\centering 
\includegraphics[width=0.49\linewidth]{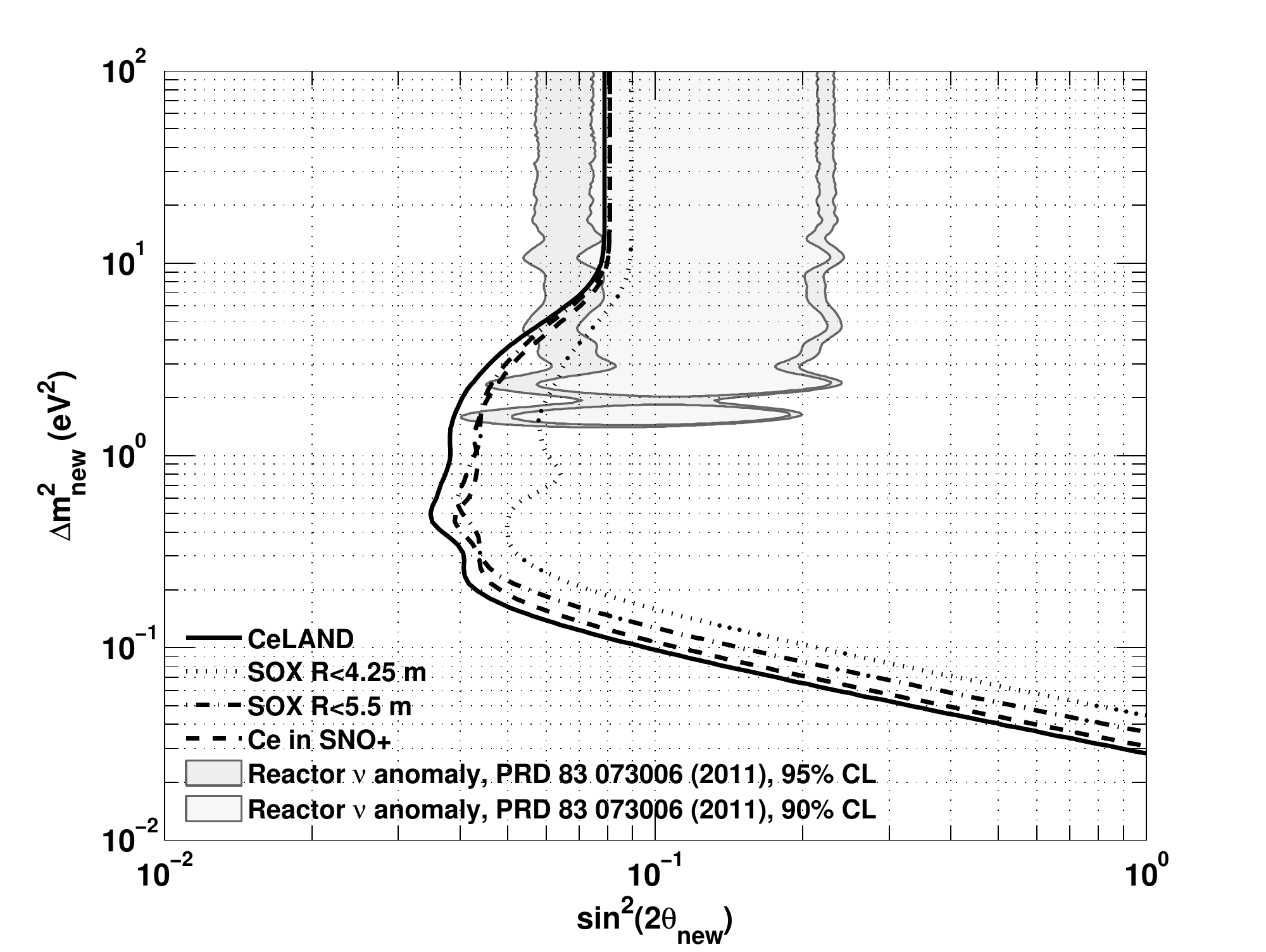}
\includegraphics[width=0.49\linewidth]{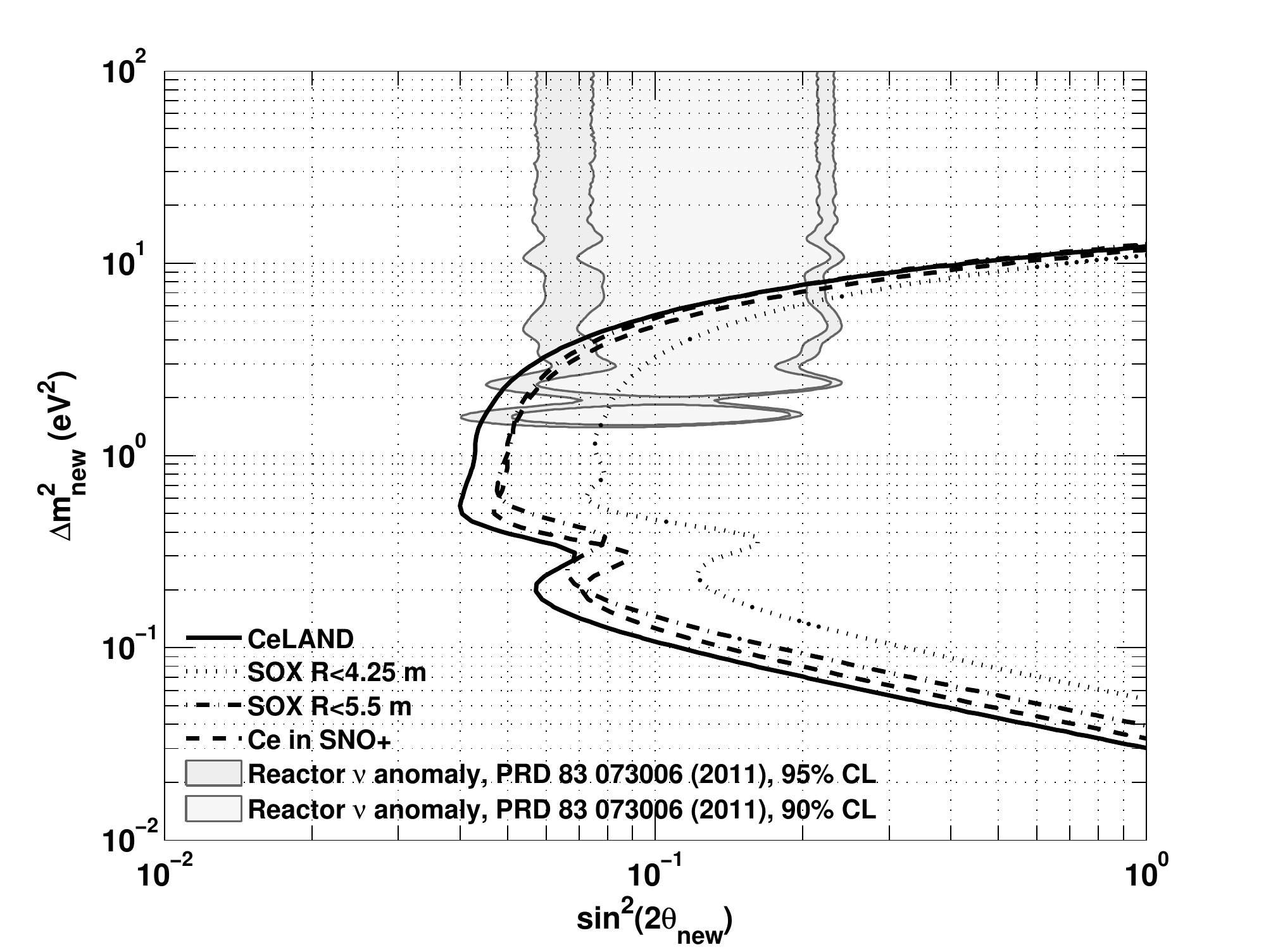}
\caption{\label{fig:detector_comparison} Sensitivity contours of a source experiment at the KamLAND, 
Borexino, and  SNO+ detectors. See figure~\ref{fig:ref} for parameters.
Left: Rate+shape sensitivity. Right: Free rate sensitivity.}
\end{figure*}

\begin{figure}[ht!]
\centering \includegraphics[width=\linewidth]{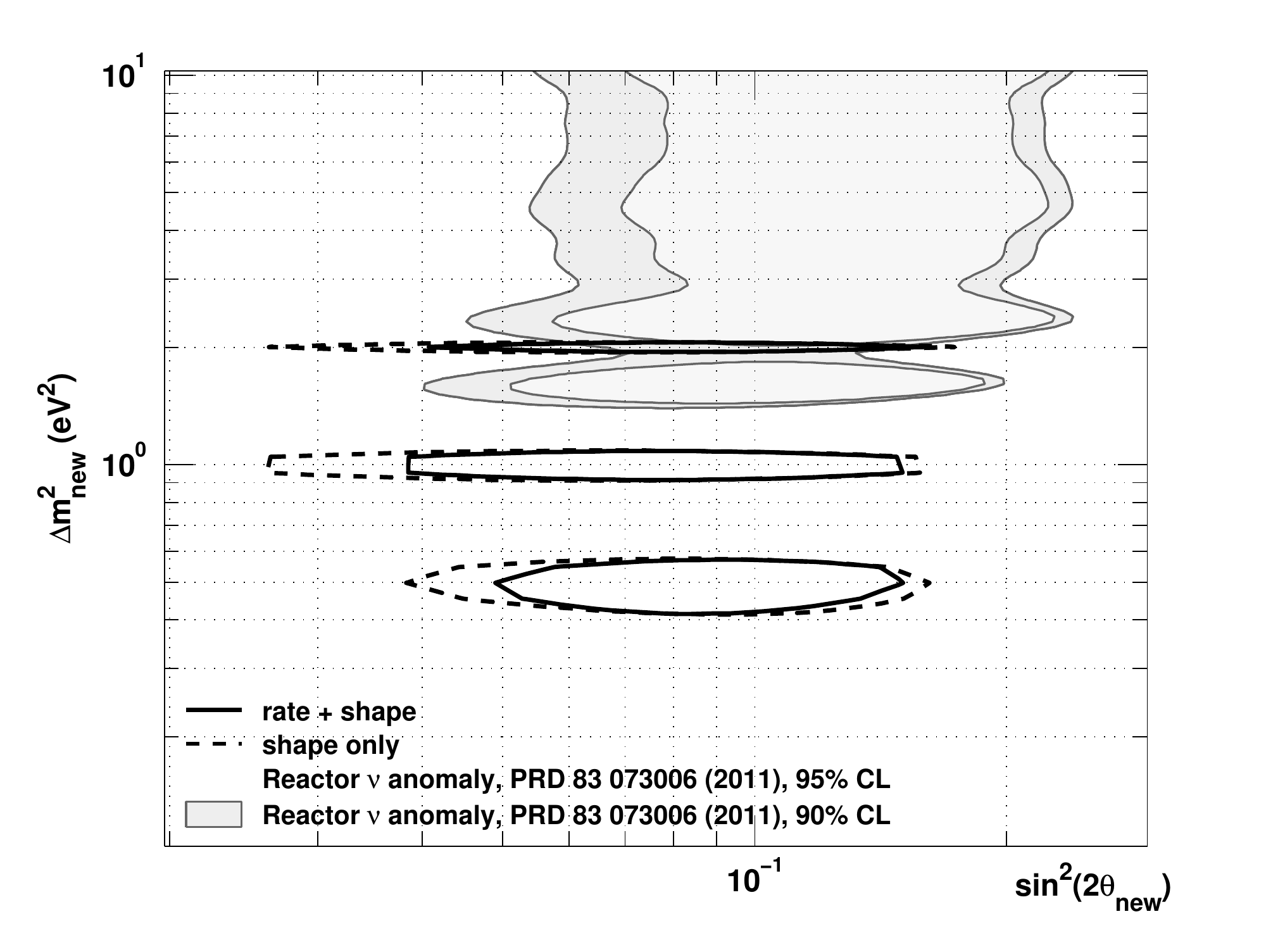}
\caption{\label{fig:potential} Illustrative contours for input oscillation signals at \Dmn{} = 0.5, 1,
\SI{2}{eV^2} and $\stn = 0.1$. Acceptance contours are given at the \SI{99}{\%} confidence level.
See figure~\ref{fig:ref} for other parameters.}
\end{figure}

After minimizing the $\chi^2$ function over the nuisance parameter $\mathrm{\alpha}$, the \SI{90}{\%},
\SI{95}{\%}, and \SI{99.73}{\%}~exclusion contours are computed as a function of \Dmn{} and \tn{} such that
$\Delta \chi^2 = \chi^2(\tn,\Dmn)-\chi^2_{min}<$ 4.6, 6.0 and 11.8, respectively. Such contours are shown on
figure~\ref{fig:ref} for both rate+shape and free rate analysis, assuming the generic experimental
configuration described above.

The shape of the sensitivity contours can be understood in the following way. At low $\Dmn \lesssim
\SI{0.1}{eV^2}$, the typical oscillation lengths are much larger than the detector size, and the \anu{}
survival probability approximates to $\mathrm{{\cal{P}}(L,E) \approx 1 - A\,\stn \times \left(\Dmn\right)^2}$,
where A is a constant. In this regime, sensitivity contours defined by constant $\Delta\chi^2$ values impose
then a linear relationship in logarithmic scale between \stn{} and \Dmn{}.

In the $\Dmn \sim \SIrange[range-phrase = -]{0.1}{5}{eV^2}$ regime, oscillation periods are smaller than the
detector size, but are also small enough so that they are slightly damped by the detector energy and vertex
resolutions (see figure~\ref{fig:SignalLoE} and related discussion). Moreover, the energy and distance binning
sizes, which are of the order of the detector energy and vertex resolutions, are small enough so that the
oscillation period is properly sampled. This \Dmn{} regime is the region where the sensitivity to short
baseline oscillations is maximum.

Finally, in the high $\Dmn{} \gtrsim \SI{10}{eV^2}$ regime, the typical oscillation periods are
much smaller than the energy and distance binning sizes, and are seriously damped by the detector energy and
vertex resolutions. Therefore, the survival probability is averaged out to a constant value, which depends
only on the \stn{} parameter. In a free rate analysis, no information on the (\Dmn{}, \tn{}) oscillation
parameters can be recovered. In a rate+shape analysis, the rate deficit can be used to infer the \stn{} mixing
parameter, leading to contours which do not depend on the squared mass splitting \Dmn{}.

The estimated sensitivities to short baseline oscillations assuming the deployment scenarios discussed for the
KamLAND, Borexino and SNO+ detectors in section~\ref{sec:neutrino_detectors} are shown on
figure~\ref{fig:detector_comparison} for both the rate+shape and free rate analysis. A \SI{1.5}{\%}
systematic uncertainty on the initial source activity has been assumed for the rate+shape contours
calculation. As expected, the deployment scenario at KamLAND offers the best sensitivity contours among all
the detectors discussed previously.

Examples of the discovery potential of a \ce{}-\pr{} ANG experiment to
short baseline neutrino oscillations, still assuming the generic experimental configuration described
previously, are illustrated on figure~\ref{fig:potential}. It shows the acceptance contours at the \SI{99}{\%} 
confidence level of the inferred oscillations parameters if one assumes \Dmn{} = 0.5, 1, \SI{2}{eV^2},
respectively.

\subsection{Impact of experimental parameters}
\subsubsection{\label{sec:par_source} Activity and associated uncertainty}

\begin{figure*}[ht!]
\centering 
\includegraphics[width=0.49\linewidth]{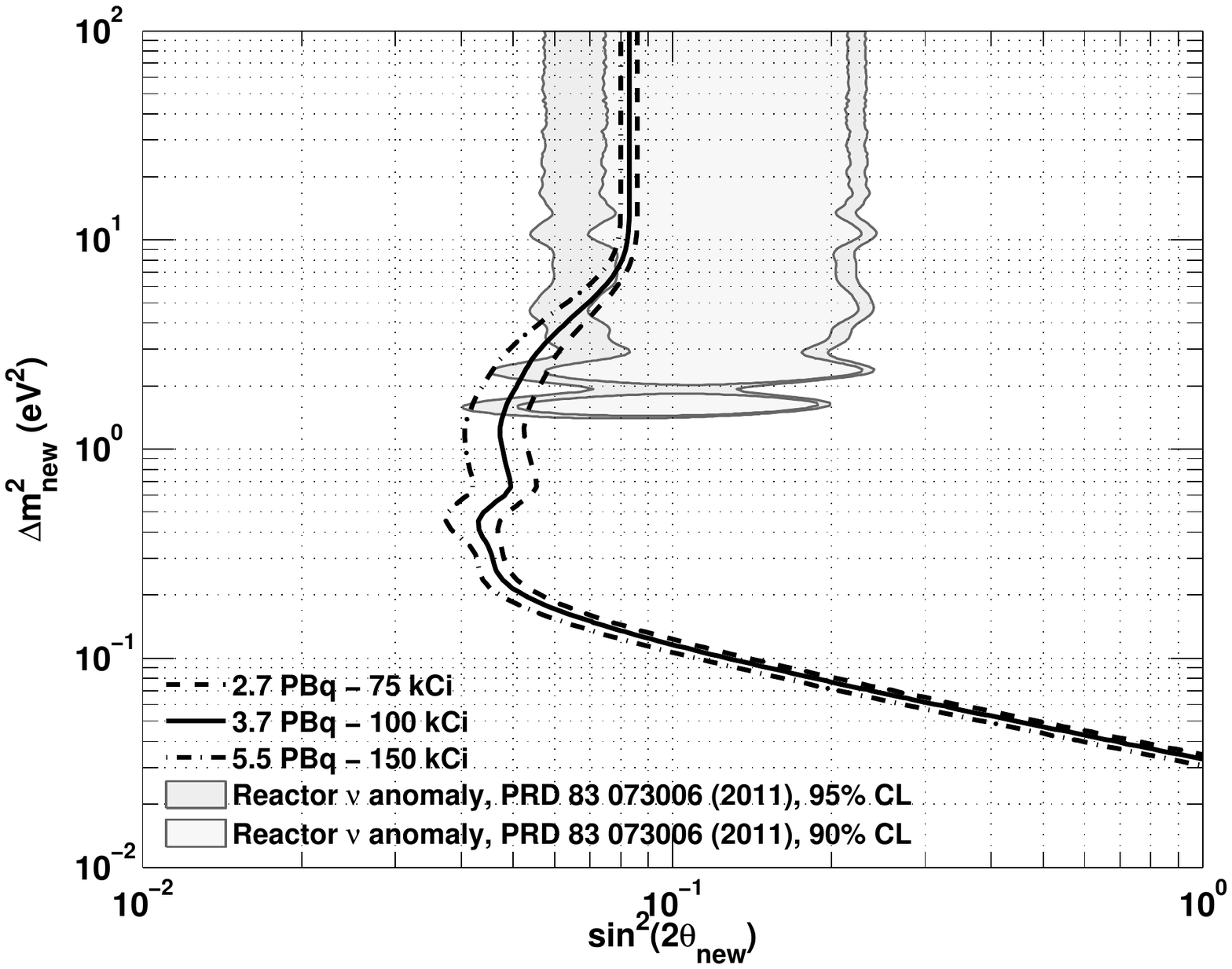}
 \includegraphics[width=0.49\linewidth]{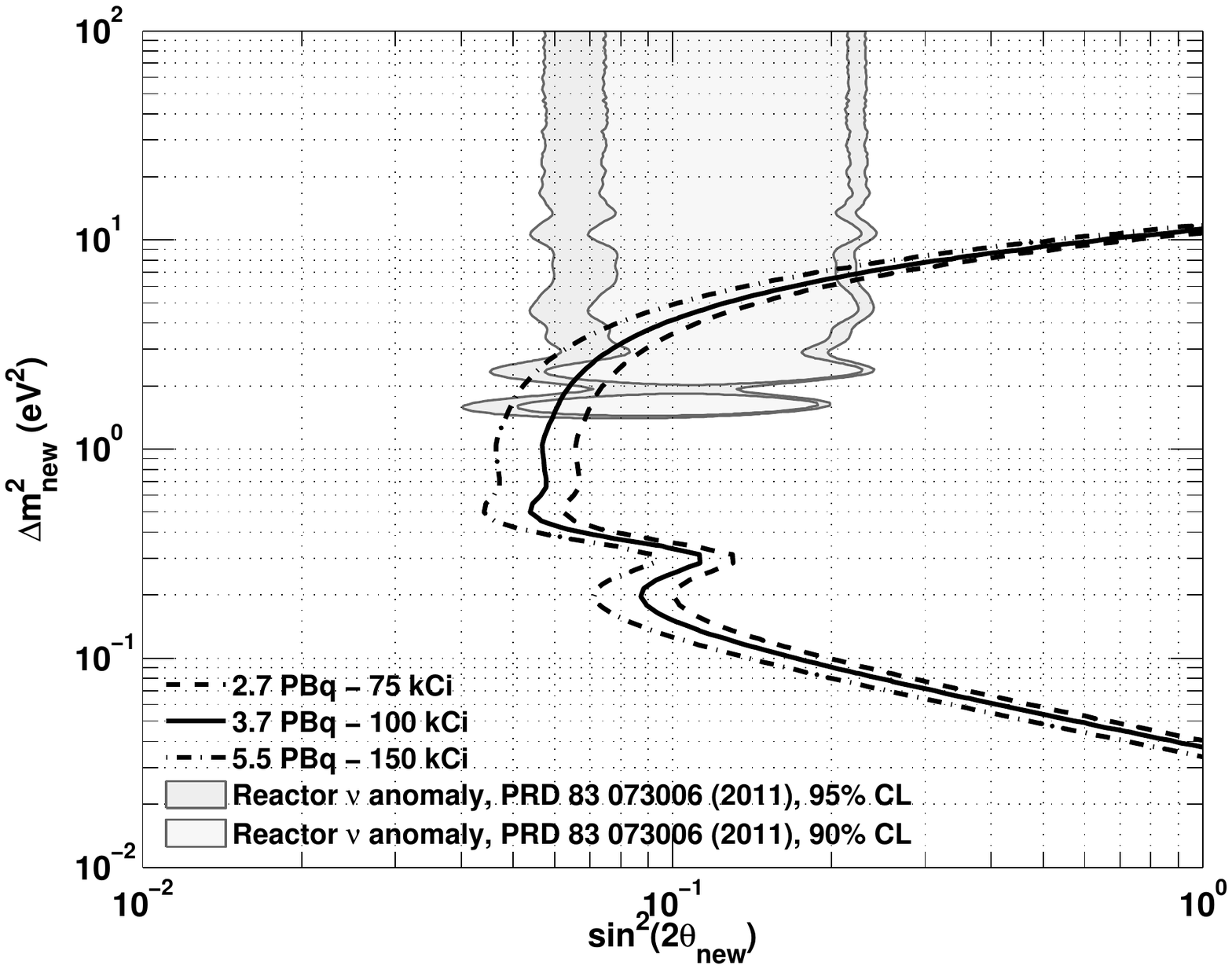}
\caption{\label{fig:activity} Influence of source activity on sensitivity, with a rate+shape analysis (left) 
and a free rate analysis (right). See figure~\ref{fig:ref} for other parameters.}
\end{figure*}

Figure \ref{fig:activity} shows how the source activity affects the sensitivity to short baseline
oscillations. Increasing the source activity has little effect in the low $\Dmn \lesssim \SI{0.1}{eV^2}$
regime. In this region the oscillation length is larger than the detector size. In this domain
of the parameter space, the $\mathrm{\chi^2}$ sensitivity study compares an overall weighted number of IBD
events between the observed and expected signals.
Therefore, because of the $\mathrm{\sigma_N}$ limiting systematic uncertainty the sensitivity contours only slightly
change with the different source activities studied here.
However, the intermediate $\SI{0.1}{eV^2} \lesssim \Dmn \lesssim \SI{10}{eV^2}$ region is
more impacted because in this regime, many oscillation lengths can be sampled in the
detector. Increasing the statistics improve the precisions to bin-to-bin changes and thus the search for
oscillation patterns. Finally, in the high $\Dmn \gtrsim \SI{10}{eV^2}$ regime, the source activity
slightly affects the experimental sensitivity with a rate+shape analysis (figure  \ref{fig:activity}, left)
the free rate analysis being insensitive in that domain (figure \ref{fig:activity}, right). The sensitivity is
here limited by the \SI{1.5}{\%} normalization uncertainty, larger than the $\sim \SI{1}{\%}$
statistical uncertainty.

\begin{figure*}[ht!]
\centering 
 \includegraphics[width=0.49\linewidth]{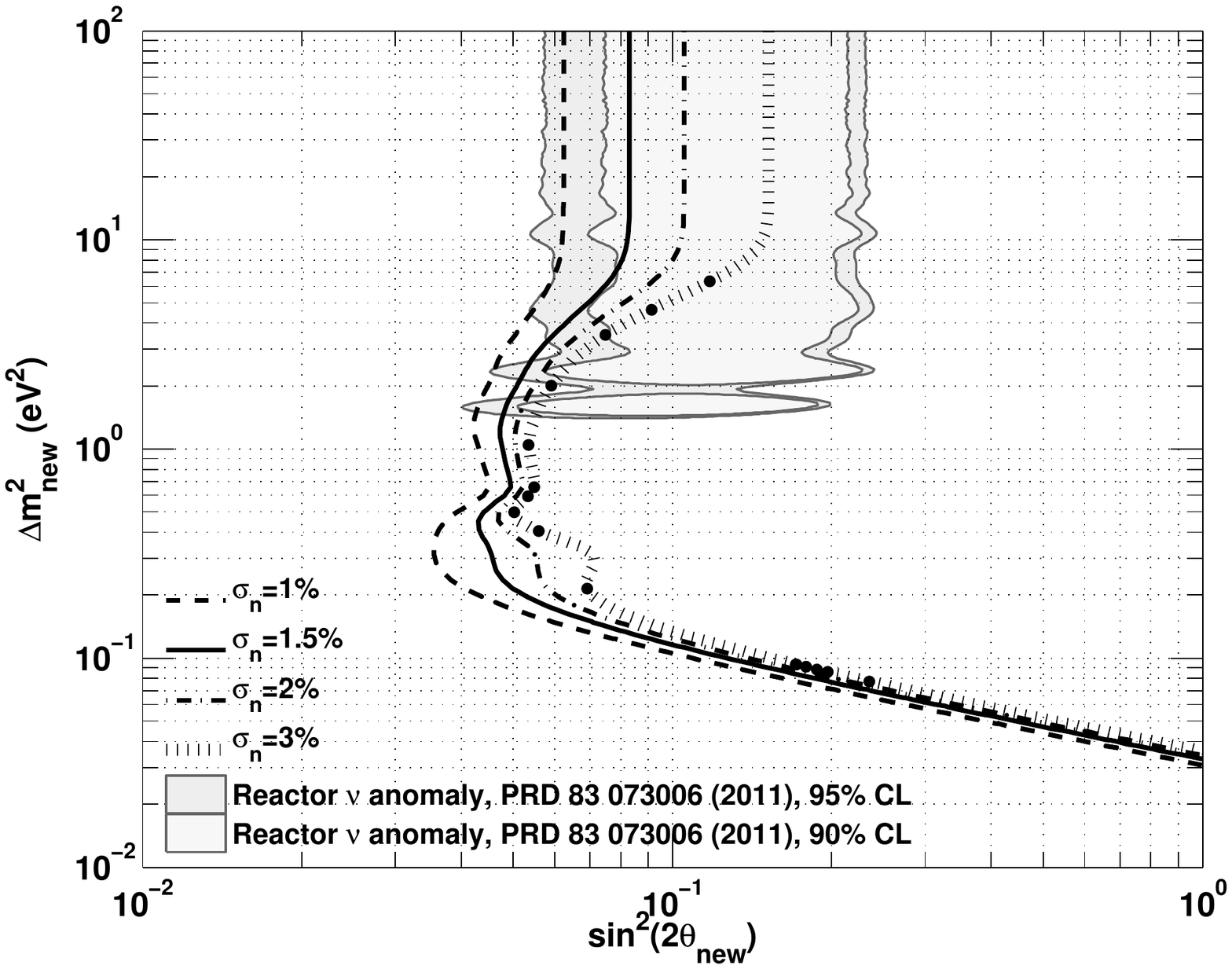}
 \includegraphics[width=0.49\linewidth]{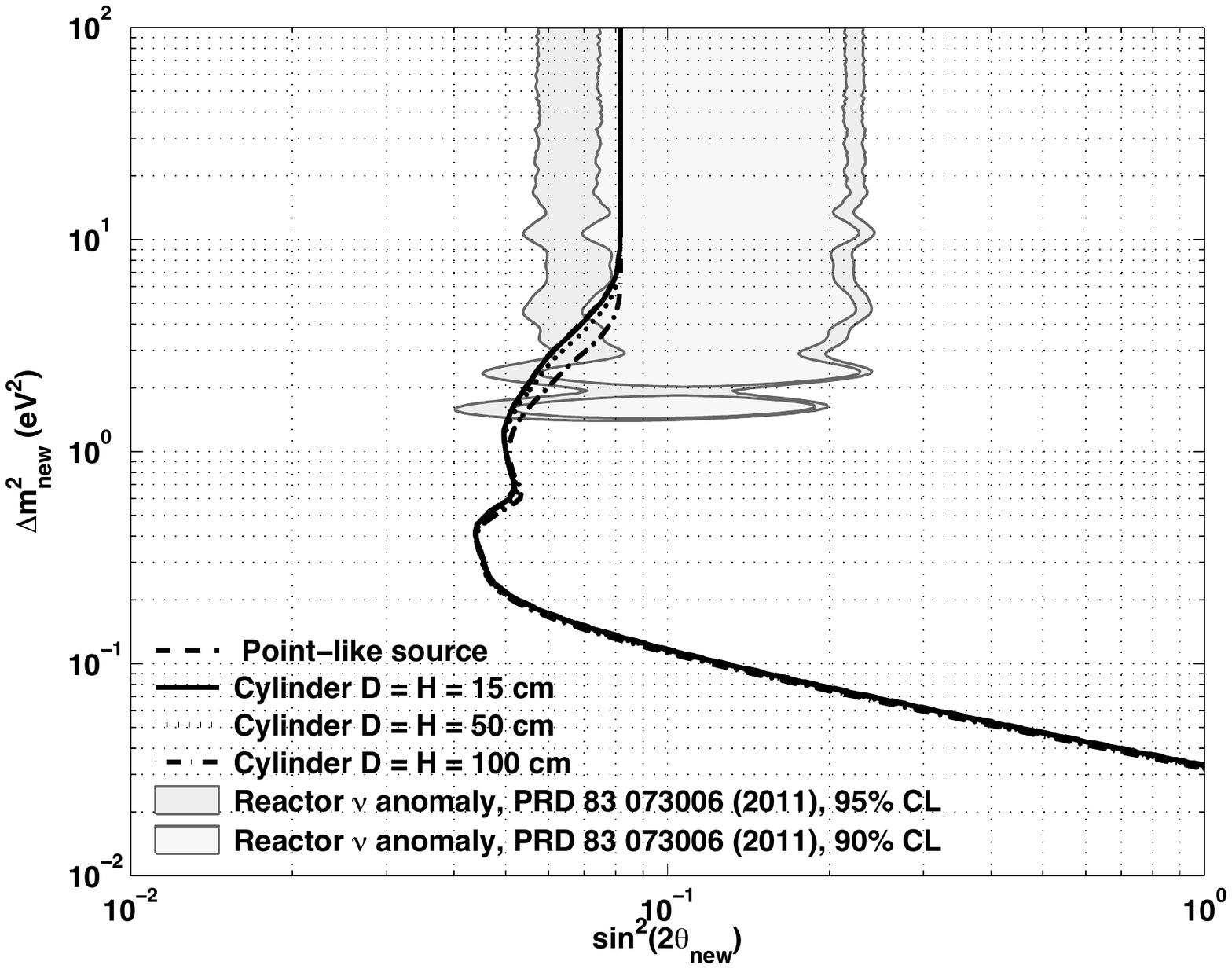}
\caption{\label{fig:normalization} Left: influence of normalization uncertainty on sensitivity.
Right: influence of source extension on sensitivity.  See figure~\ref{fig:ref} for other parameters.}
\end{figure*}

Left panel of figure \ref{fig:normalization} illustrates the impact of the source activity knowledge, $\mathrm{\sigma_N}$,
to the $\mathrm{\chi^2}$ sensitivity contours. As previously stated, uncertainties on the source activity dominate
over statistical uncertainties in the high $\Dmn{} \gtrsim \SI{10}{eV^2}$ regime. Degrading the
activity uncertainties from \SI{1}{\%} down to \SI{3}{\%} has therefore a strong impact on the sensitivity
contours in this region. The lowest part of the sensitivity contours ($\Dmn \lesssim
\SI{0.5}{eV^2}$) is also affected because of the measurement degeneracy between \Dmn{} and
\stn{} in this region where the sensitivity only depends on an overall weighting of events.
A better knowledge on source's activity improves the global IBD events counting uncertainty and directly results in a better
determination of the oscillation parameters. The $\mathrm{\chi^2}$ contours are also impacted to a lesser extent in the
intermediate \Dmn{} region. Such a study
demonstrates that measuring the source activity with a percent level precision is therefore a prime
requirement in order to ensure a good overall sensitivity.

\begin{figure*}[ht!]
\centering 
\includegraphics[width=0.49\linewidth]{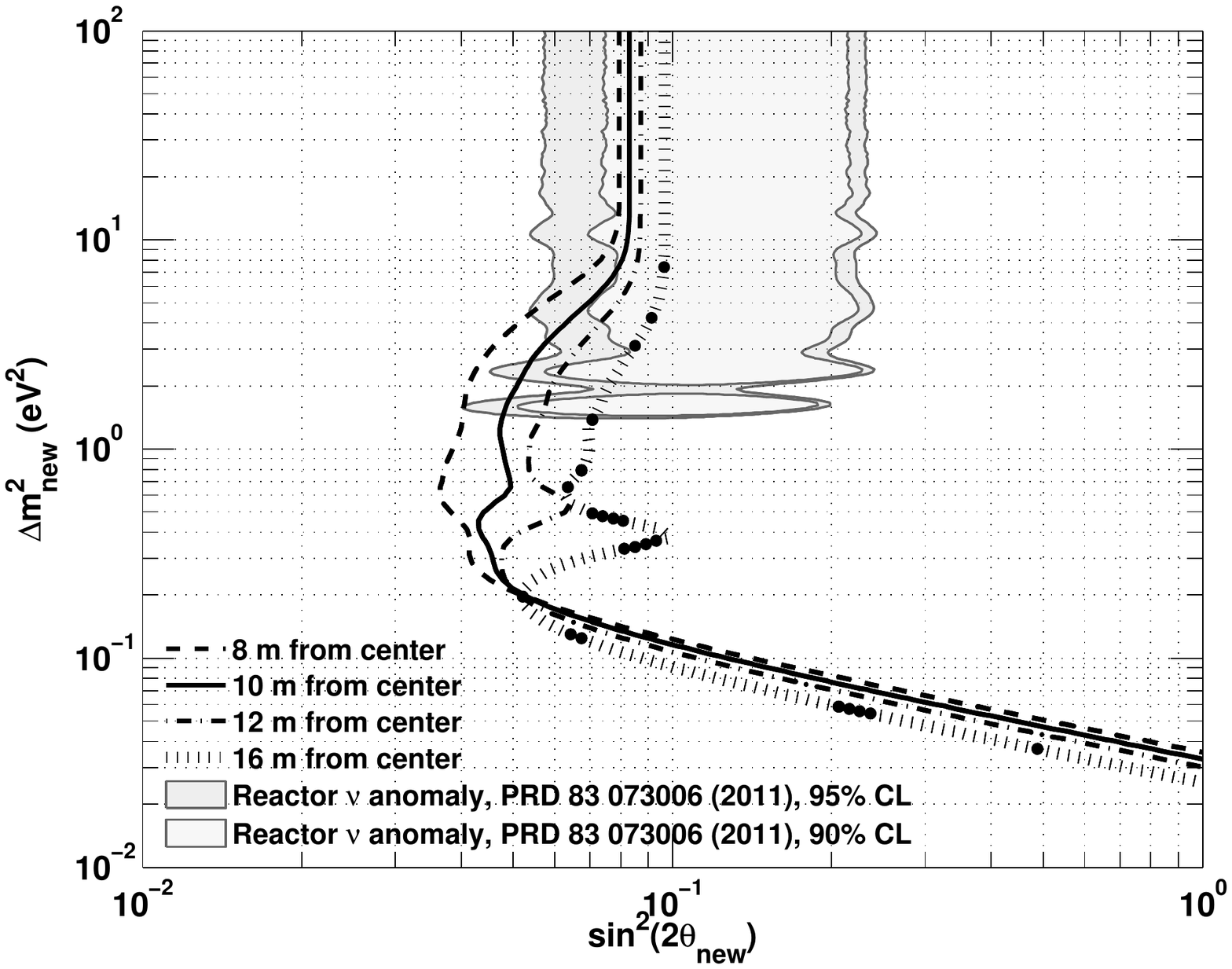}
\includegraphics[width=0.49\linewidth]{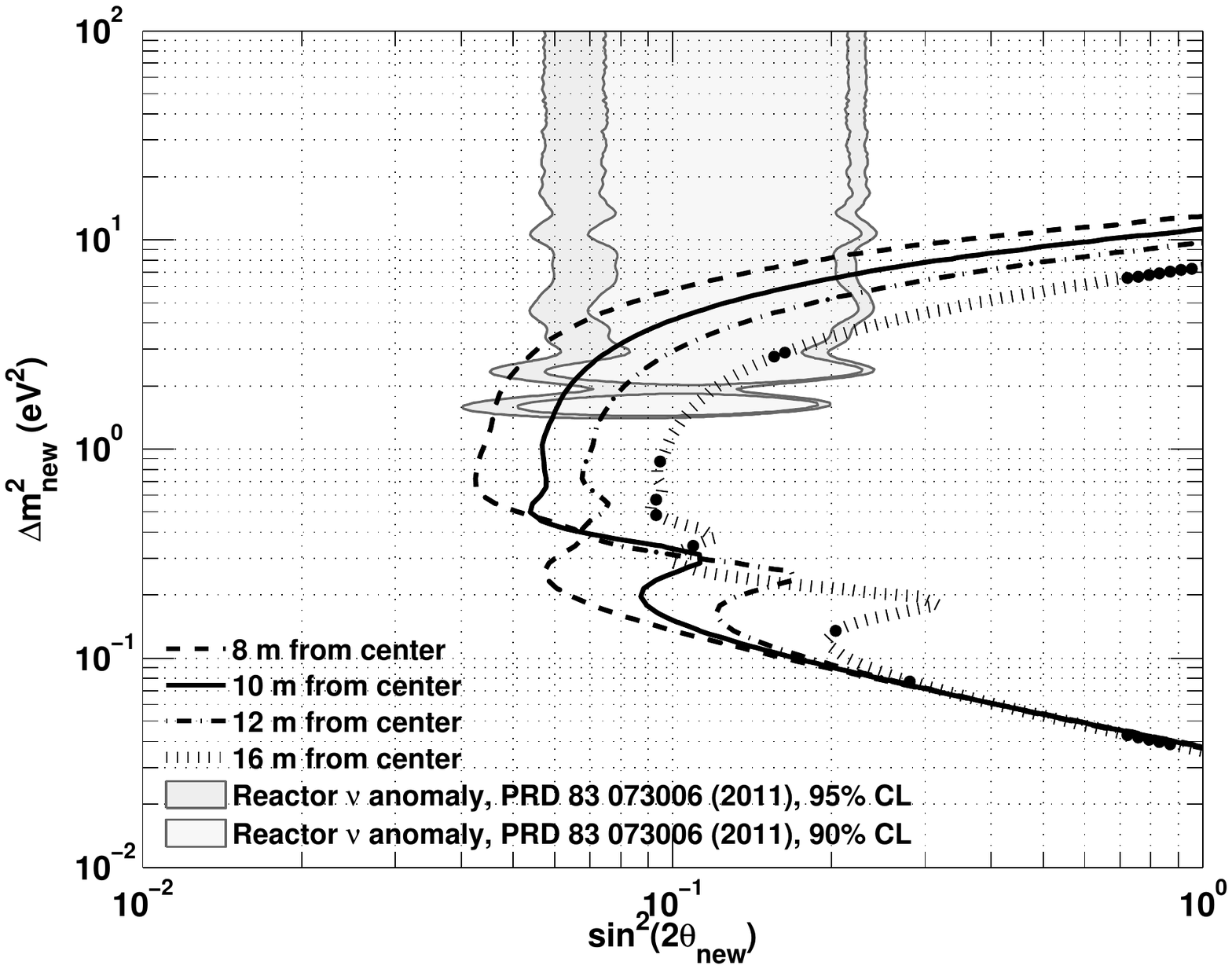}
\caption{\label{fig:distance} Influence of distance between source and detector center on
sensitivity, with a rate+shape analysis (left) and a free rate analysis (right). In the
extreme case the overall statistics vary from \SI{6.2e2}{events} to \SI{2.7e4}{events}. See 
figure~\ref{fig:ref} for other parameters.}
\end{figure*}

\subsubsection{Source spatial extension}
Up to now, the L and E binned spectra modeling presented in section \ref{sec:signal} assumed the \ce{}-\pr{}
ANG to be a point-like source. At the end of the manufacturing process, the \ce{}-\pr{} source will be packed
in a cylindrical double capsule made of stainless steel, with equal diameter and height of roughly
\SI{15}{cm}. This section investigates the effect of a spatially extended source on the sensitivity contours
to short baseline oscillations.

For a spatially extended source, the modeling of the binned L and E spectra is done averaging equation
\ref{eq:nu_rate_ext} on the source volume. This is done through a Monte Carlo simulation by randomly drawing many
point-sources in the cylindrical source volume and averaging the corresponding L and E spectra. Right panel of
figure \ref{fig:normalization} shows the impact of the source spatial extension on the sensitivity contours,
for cylinders of equal diameters and heights. Averaging equation \ref{eq:nu_rate_ext} over the source volume
makes the amplitude of the \anu{} oscillations damped with respect to equation \ref{eq:survival_proba}, while
keeping the same mean deficit. Therefore, sensitivity to short baseline oscillations is especially lost in the
intermediate \Dmn{} regime compared to the point-like source case. Sensitivity remains
almost unaffected in the low \Dmn{} regime because, once again, the oscillation length is
larger than the detector size and makes the detector insensitive to any oscillation damping. In the high
\Dmn{} region, contours are also unaffected by the source spatial extension. In this
regime, the oscillation pattern is seriously damped by the detector vertex and energy resolution, making the
sensitivity to short baseline oscillations being entirely driven by a rate deficit.

As further shown by right panel of figure \ref{fig:normalization}, the spatial extension of a \ce{}-\pr{}
source with H=D=\SI{15}{cm} such as the source provided by \ma{} has a negligible impact on the sensitivity
contours with respect to the point-like source case. Indeed, an oscillation half-length of \SI{15}{cm} or
lower corresponds to \Dmn{} values above \SI{20}{eV^2} where the detector energy and
distance resolutions have already damped the oscillation patterns. A source with spatial extension larger than
the detector vertex resolution would be necessary to significantly degrade the sensitivity contours. A
\SI{1}{m} half oscillation length corresponds to \Dmn{} around \SI{3}{eV^2} where
logically the sensitivity is the most degraded with an hypothetical \SI{1}{m} sized source.

\begin{figure*}[ht]
\centering 
\includegraphics[width=0.49\linewidth]{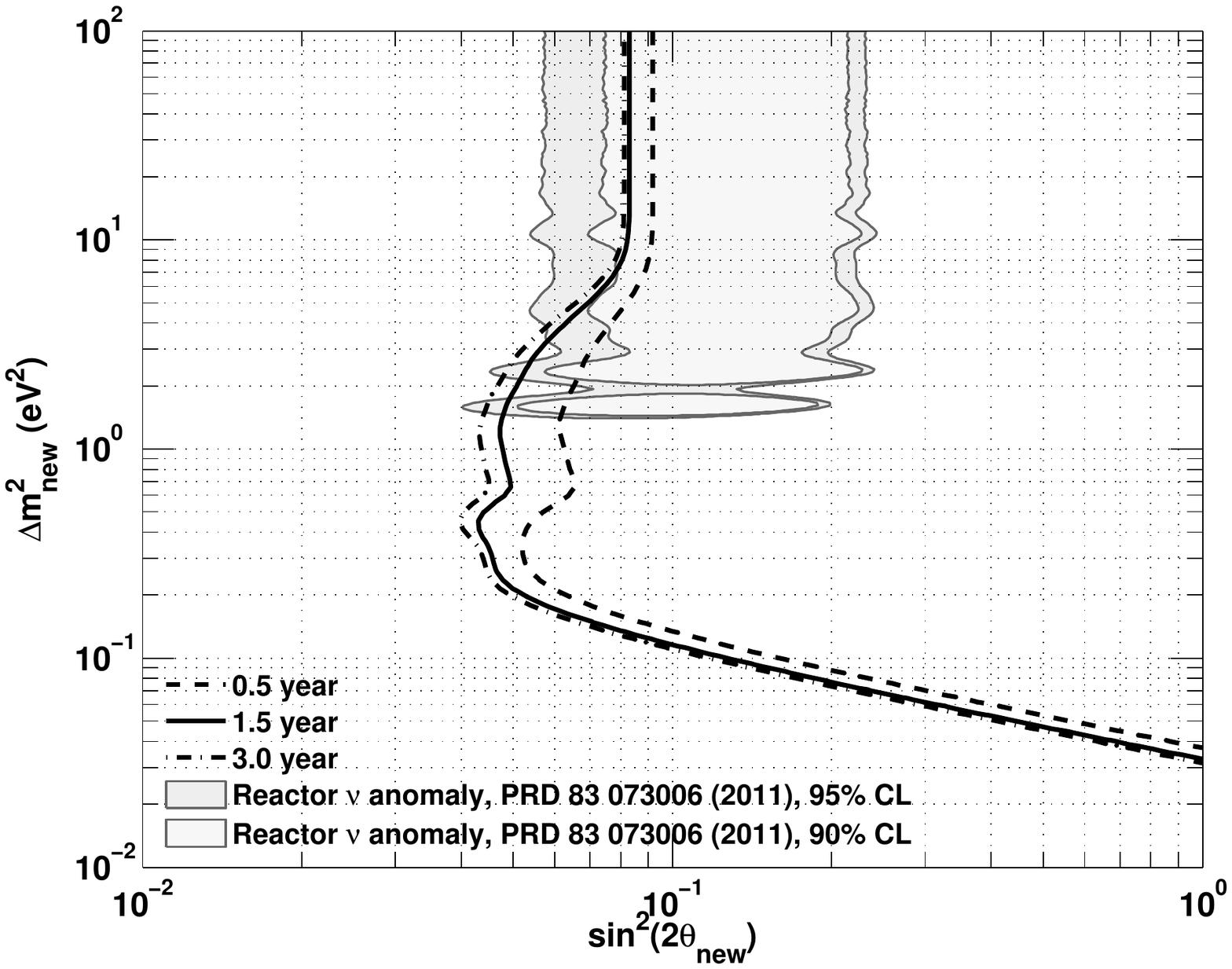}
\includegraphics[width=0.49\linewidth]{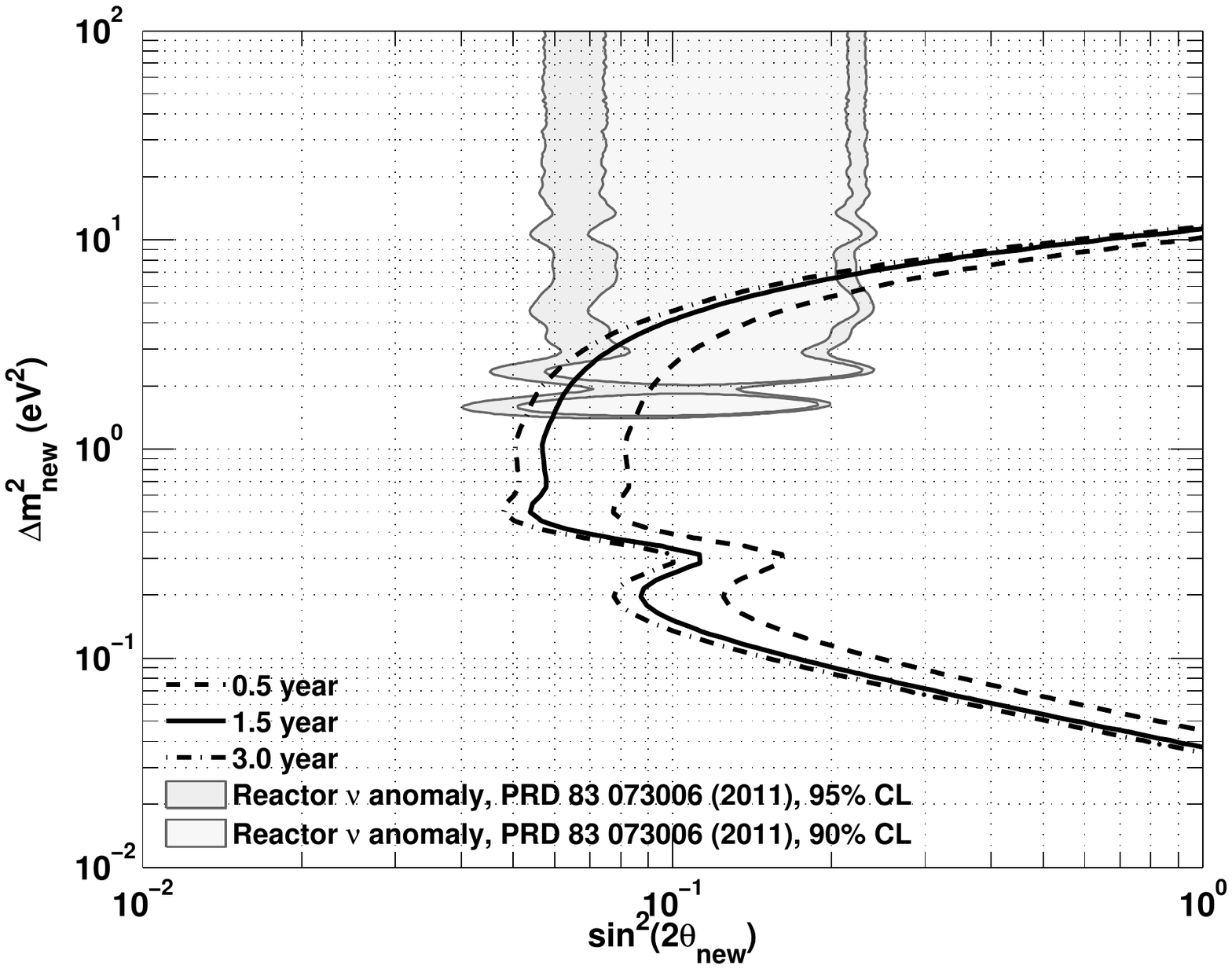}
\caption{\label{fig:duration} Influence of data taking time on sensitivity, 
with a rate+shape analysis (left) and a free rate analysis (right). See figure~\ref{fig:ref} for other 
parameters.}
\end{figure*}

\subsubsection{Source deployment location}
The three detectors suitable for a \anu{} source experiment at short baselines (see section
\ref{sec:neutrino_detectors}) have quite a similar design. Any difference in the experimental configuration
offered by such detectors arise then from the source deployment location with respect to the detector center.
The deployment location directly determines the distance between the ANG and the detector center, and
therefore inversely scales quadratically the statistical uncertainties in the expected number of \anu{} events
both as a function of energy $E$ and baseline $L$. Therefore, the resulting effect on the sensitivity contours
to short baseline oscillations is expected to be similar than the impact of the ANG activity.

Figure~\ref{fig:distance} shows the sensitivity contours for different source to detector distances. Both the
rate+shape and free rate analysis are strongly impacted. The change in the sensitivity contours at low and
high \Dmn{} is mostly due to the change in the statistical uncertainties caused by the
source-detector distance, because oscillations are unresolved in these regimes. However, the impact of the
source-detector distance is quite different from the impact of the source activity in the $\SI{0.1}{eV^2}
\lesssim \Dmn \lesssim \SI{0.5}{eV^2}$ region. The ``turnover point'' between the low \Dmn{} and intermediate
\Dmn{} regimes is reached when a first oscillation maximum starts to be fully contained and sampled by the
detector. This situation depends both on the source location with respect to the detector center and the
detector size. As shown by right panel of figure~\ref{fig:distance}, it corresponds to $\Dmn = \SI{0.2}{eV^2}$
and $\stn = 0.09$ for the generic experimental configuration described previously (i.e.
an ANG located \SI{10}{m} away from the center of a \SI{5.5}{m} radius spherical detector). Detectors placed
further away than \SI{10}{m} from an ANG present ``turnover points'' in a free
rate analysis which correspond to lower \Dmn{} (i.e. larger oscillation periods) and
lower \stn{} values. This is first because of the resulting decrease in the
statistical uncertainties, and second because larger source-detector baselines enable detectors to be
positioned on the maximum of oscillation patterns with larger periods.

From the event statistics point of view, the closer the ANG to the detector center the better. However, the
smallest source-detector baseline might not be the optimal baseline if the source-induced gamma and neutron
radiations can make a significant level of background into the detector. This concern is especially important
for the deployment of an ANG within a low background liquid scintillator detector such as those discussed in
section~\ref{sec:neutrino_detectors}.

The statistics is also enhanced by data taking time. However, this effect is attenuated by the decrease in
activity with time, as shown by figures~\ref{fig:duration}. For our generic reference experiment the expected
number of events are \num{8.0e3}, \num{1.6e4}, and \num{2.1e5} for 0.5, 1.5, and \SI{3}{years} of data taking
respectively, assuming a \SI{100}{\%} detection efficiency.
Therefore, the deployment of the \ce{}-\pr{} ANG for \SI{1.5}{years} is a good compromise, the sensitivity
being marginally improved for longer data taking times.

\subsubsection{Detector energy and vertex resolutions}
\begin{figure*}[ht!]
\centering 
\includegraphics[width=0.49\linewidth]{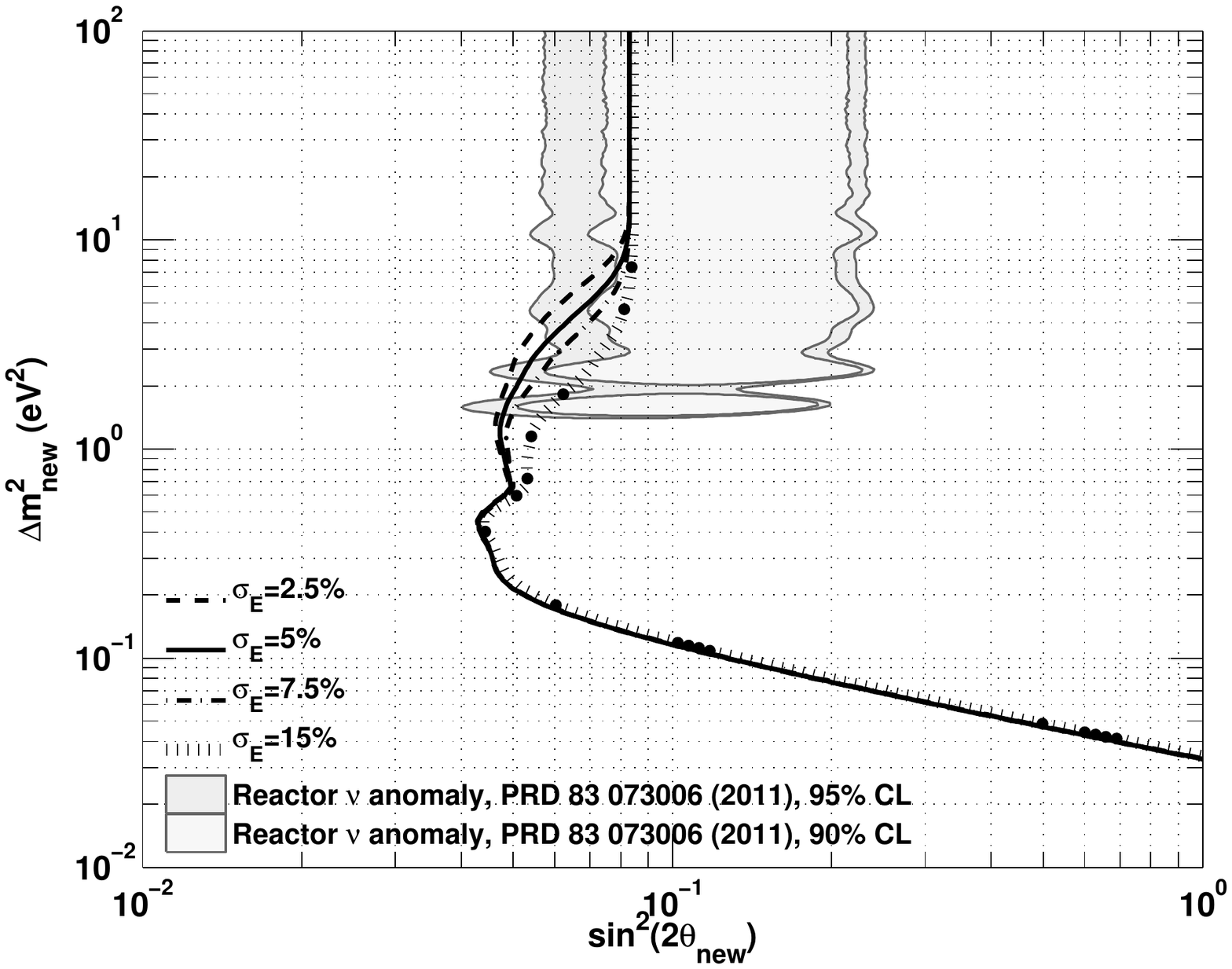}
 \includegraphics[width=0.49\linewidth]{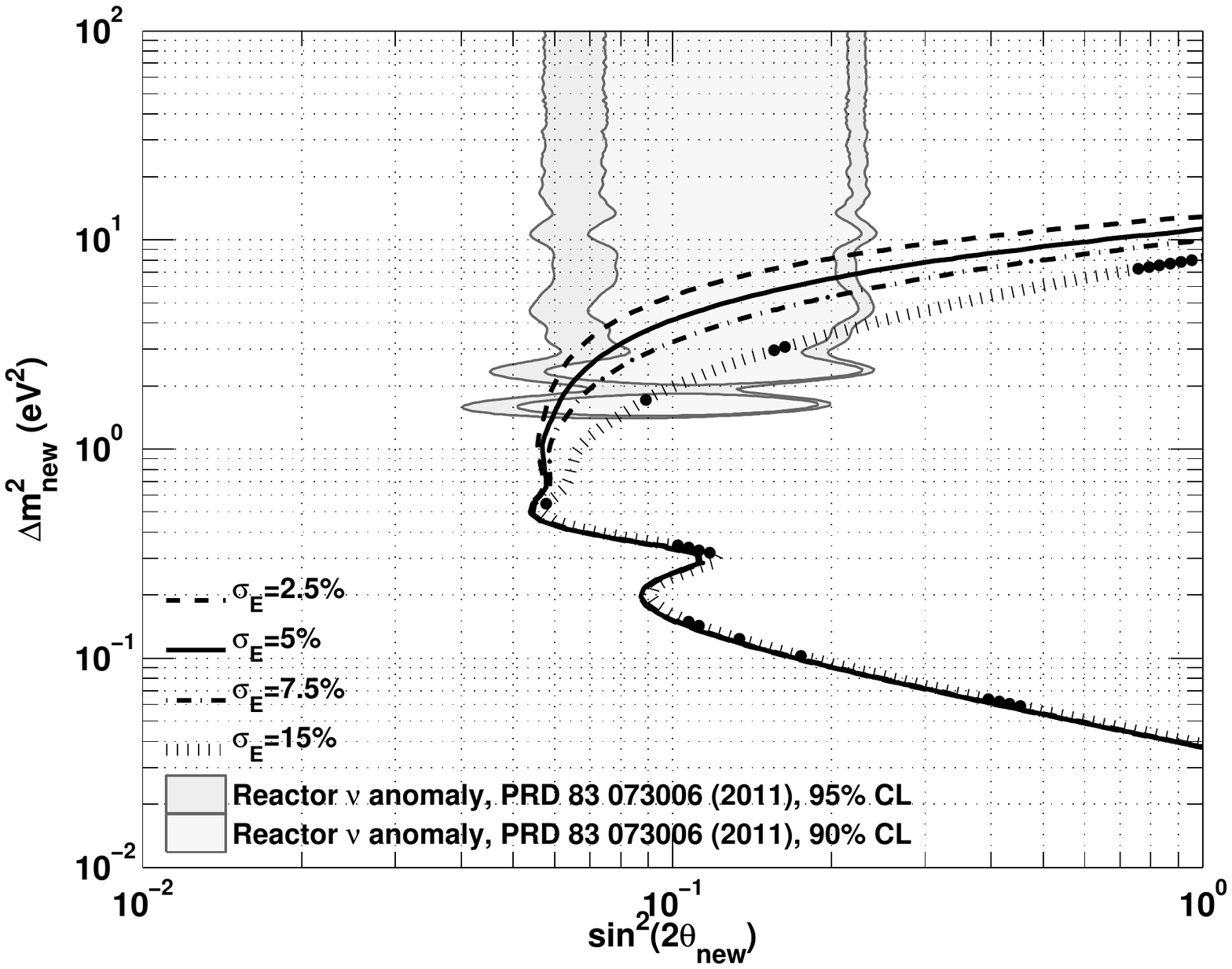}
\caption{\label{fig:energy_reso} Influence of energy resolution on sensitivity, with a rate+shape analysis 
(left) and a free rate analysis (right). See figure~\ref{fig:ref} for other parameters.}
\end{figure*}

\begin{figure*}[ht]
\centering 
\includegraphics[width=0.49\linewidth]{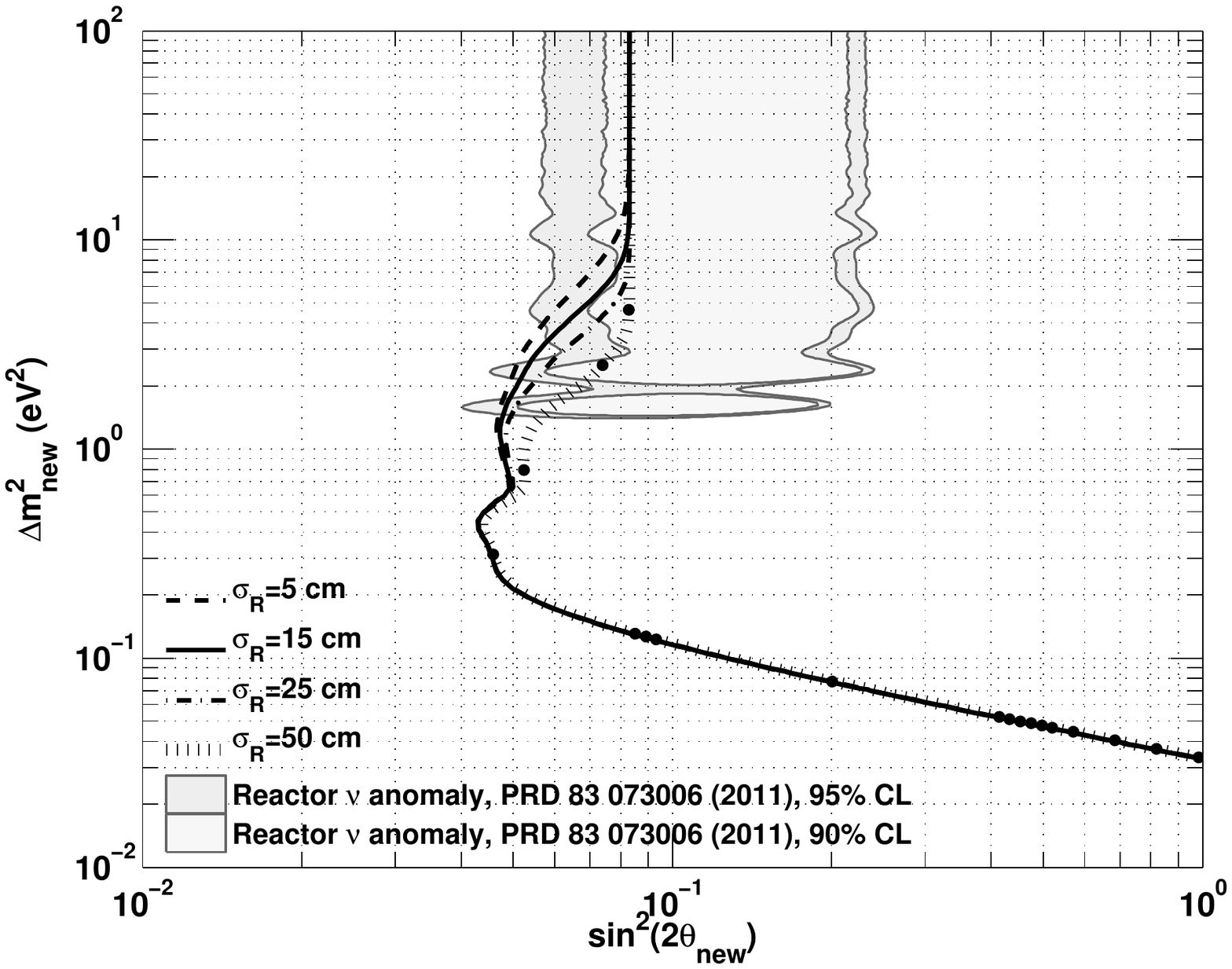}
\includegraphics[width=0.49\linewidth]{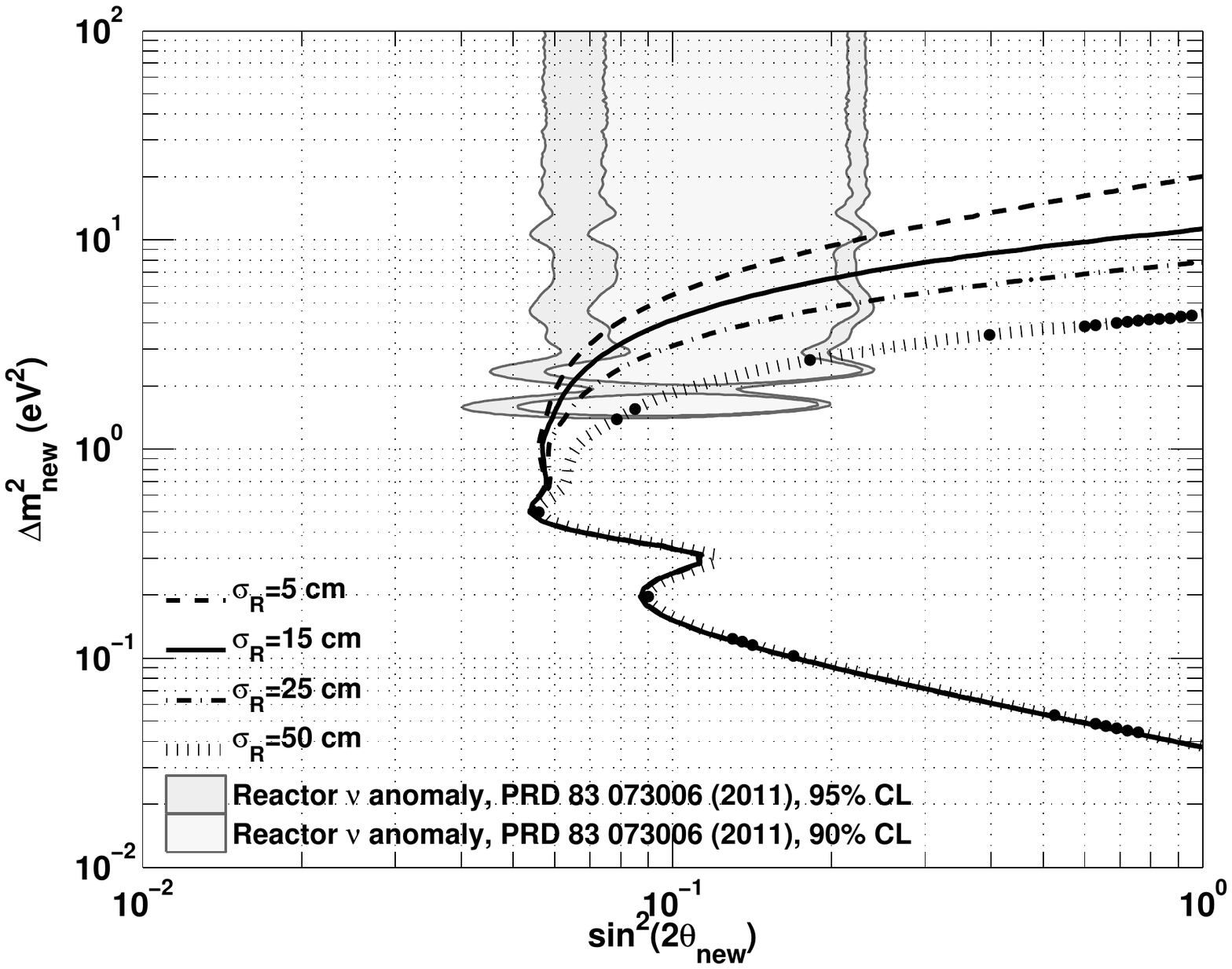}
\caption{\label{fig:vertex_reso} Influence of vertex resolution on sensitivity, with a rate+shape analysis 
(left) and a free rate analysis (right). See figure~\ref{fig:ref} for other parameters.}
\end{figure*}

The effect of varying the detector energy and vertex resolutions on the sensitivity contours is shown on
figure~\ref{fig:energy_reso} and figure~\ref{fig:vertex_reso}, respectively. The energy resolution of the
generic detector considered in the present study was varied from \SI{2.5}{\%} to \SI{15}{\%} and was assumed
to be independent of energy. The vertex resolution was varied from 5 to \SI{50}{cm}. The change in the
sensitivity contours due to the degradation of energy and vertex resolution is similar, and occurs only in the
$\Dmn \gtrsim \SI{1}{eV^2}$ region for a free rate analysis. As explained in
section~\ref{sec:signal}, and as shown on figure~\ref{fig:SignalLoE}, the oscillation pattern is significantly
washed out by the finite detector energy and vertex resolution for $\Dmn \gtrsim \SI{1}{eV^2}$. Concerning the
rate+shape analysis, the sensitivity contours remains unchanged for $\Dmn \gtrsim \SI{10}{eV^2}$ because in
this region, the experiment is only sensitive to a rate deficit.

The three detectors discussed in section~\ref{sec:neutrino_detectors} typically have energy resolutions and
vertex resolutions of $\mathrm{\Eres{}/\sqrt{E_{vis}(MeV)}}$ and $\mathrm{\Lres{}/\sqrt{E_{vis}(MeV)}}$,
respectively. As shown by figure~\ref{fig:energy_reso} and~\ref{fig:vertex_reso}, such detectors then present
good enough vertex and energy reconstruction performances for a short baseline ANG experiment. However, calibration
of the detector response to vertex and energy reconstruction has to be carefully done through the full
detection volume, especially to pinpoint and correct for any position dependency.

\section{\label{sec:backgrounds}Backgrounds}
As already stated in section~\ref{sec:neutrino_detectors}, the backgrounds to \anu{} detection in the KamLAND,
Borexino and SNO+ detectors are negligible, thanks to large overburdens and low radioactivity materials.
However, radioactive contaminants left in the ANG after the manufacturing process can bring several
backgrounds to the detector, divided into gamma and neutron contributions. This section discusses these
possible source-induced backgrounds. A detailed simulation of the source-induced gamma and neutron backgrounds
is presented and used to estimate the rate of backgrounds in the generic experimental configuration described
previously (see section~\ref{sec:sensitivity}) as a function of the level of radioactive contaminants in the
\ce{}-\pr{} source. 

\subsection{\label{sec:Tripoli}TRIPOLI-4\textregistered{} simulation code}
Simulating the source-induced background is challenging, especially because the transportation of gamma and
neutrons through media totalling a $\mathrm{\gtrsim 10^{15}}$ attenuation factor up to the detection volume is
nearly impossible with brute force Monte-Carlo simulation (for example standard Geant4).
The TRIPOLI-4\textregistered{} Monte-Carlo simulation code was then used~\cite{BRUN}. This nuclear reactor
physics code was designed to study criticality and to transport efficiently both neutrons
and gammas with accurate models, based on databases of point-like cross sections. It has been validated by
numerous comparison with experiment and is currently used by several French nuclear companies.

TRIPOLI-4\textregistered{} was chosen for two reasons: first it offers state-of-the-art neutron transport
modeling, and second it incorporates different variance reduction techniques to deal with
the high attenuation factors of both neutrons and gammas transport in realistic configurations~\cite{BRUN}.
In particular, we used INIPOND, a special built-in module of TRIPOLI-4\textregistered{} based on the
exponential transform method~\cite{CLARK}. This exponential biasing is performed using an automatic
pre-calculation of an importance map. The importance map provides information on the probability, for each
point of the phase space, for a particle to reach the detector. It is calculated with a
simplified deterministic solver, and can be used
to calculate all possible particle paths over a virtual mesh of the geometry, together with
their associated probabilities. It allows the code to adjust the weight of a particle travelling along a given
path, in such a way that the mean quantities calculated by the code are unbiased while their variance is
reduced.
For example, below a given weight value, the particle can be killed (typically particle moving downward in our
generic reference experiment). This process allows a relatively large amount of particles to reach the target
volume, such that it is possible to obtain statistically significant results. As shown in section~\ref{sec:gamma_background},
statistical uncertainty around \SI{5}{\%} can be reached for attenuation as high as \num{e-18} with only
\num{e9} events.

But on top of the statistical uncertainty, large systematic uncertainties must be considered. The 
attenuation is so important that any bias on cross-section, density or attenuation coefficient can have a 
significant impact on the final results. Moreover, several parameters of the simulation have large 
uncertainties, such as the cerium oxide density in the source capsule. Attenuation coefficients and the 
different media density uncertainties result in an additional 10 to \SI{20}{\%} systematic uncertainty. A 
conservative \SI{50}{\%} systematic uncertainty is then affected to the simulation results presented in the 
next two sections. The quoted statistical uncertainties are an estimator of the convergence of the simulation.

Another drawback of the exponential biasing technique is that it prevents any track-by-track analysis, as can
be usually done for example in Geant4. Only predefined integral quantities, the so-called scores, are
available, such as currents crossing a surface, fluxes, reaction rates and energy deposition estimators (such
as for example the energy of the incident particle if an energy deposition occurs). The true energy deposition
is not available because the particle can survive an interaction with a lower weight, instead of releasing all
its energy. The exponential biasing method has been compared to the natural simulation (also called analog
transport) on a thinner geometry where the natural simulation could reach significant statistics. All
simulation results were found to be in very good agreement.

\subsection{\label{sec:gamma_background}Source induced gamma background}
Decay of \ce{} and \pr{} isotopes along with decay of radioactive impurities could be a source of gamma background
for the experiment, especially if the source is close to the target volume. Such gamma
backgrounds arise from the deexcitation of unstable daughter nuclei or bremsstrahlung of $\upbeta{}$
particles. The most serious source of gamma background is the \SI{2.185}{MeV} deexcitation gamma ray following
the $\upbeta^-$ decay of \pr{}, with an intensity of \SI{0.7}{\%}.

Still, other $\upgamma$-rays and X-rays can follow either from bremsstrahlung of $\upbeta{}$ particles or
other deexcitation modes of \pr{}, \ce{}, or lanthanide and actinide contaminants. Attenuation to
$\upgamma$-rays reaches a minimum of \SI{4.3e-2}{cm^2/g} around \SI{3.7}{MeV} in tungsten, a value which is
only \SI{7}{\%} less than the $\upgamma$ attenuation in tungsten at \SI{2.185}{MeV}~\cite{NIST-XCOM}.
Therefore, any $\upgamma$-rays and X-rays will be easily shielded as well.
In other words, the \pr{} \SI{2.185}{MeV} gamma emission drives the shielding thickness despite any realistic
hypothesis on the source contamination by other radioisotopes and the source induced gamma background can be
safely narrowed to the study of the \pr{} \SI{2.185}{MeV} gamma ray.

The study of the \pr{} \SI{2.185}{MeV} gamma ray escape and transport to the generic experiment target volume
is performed by generating monoenergetic gamma rays homogeneously and isotropically in the cerium oxide
source. The use of the TRIPOLI-4\textregistered{} exponential biasing method discussed in
section~\ref{sec:Tripoli} allowed to reach significant results generating only \num{e9} $\upgamma$. The
simulation shows statistical errors bars around \SI{2}{\%} per bin, as high as \SI{5}{\%} for the
\SIrange{2.1}{2.2}{MeV} bins. These statistical uncertainties indicates a reasonable convergence of the
simulation.

\begin{figure*}[ht]
\centering
\includegraphics[width=0.49\linewidth]{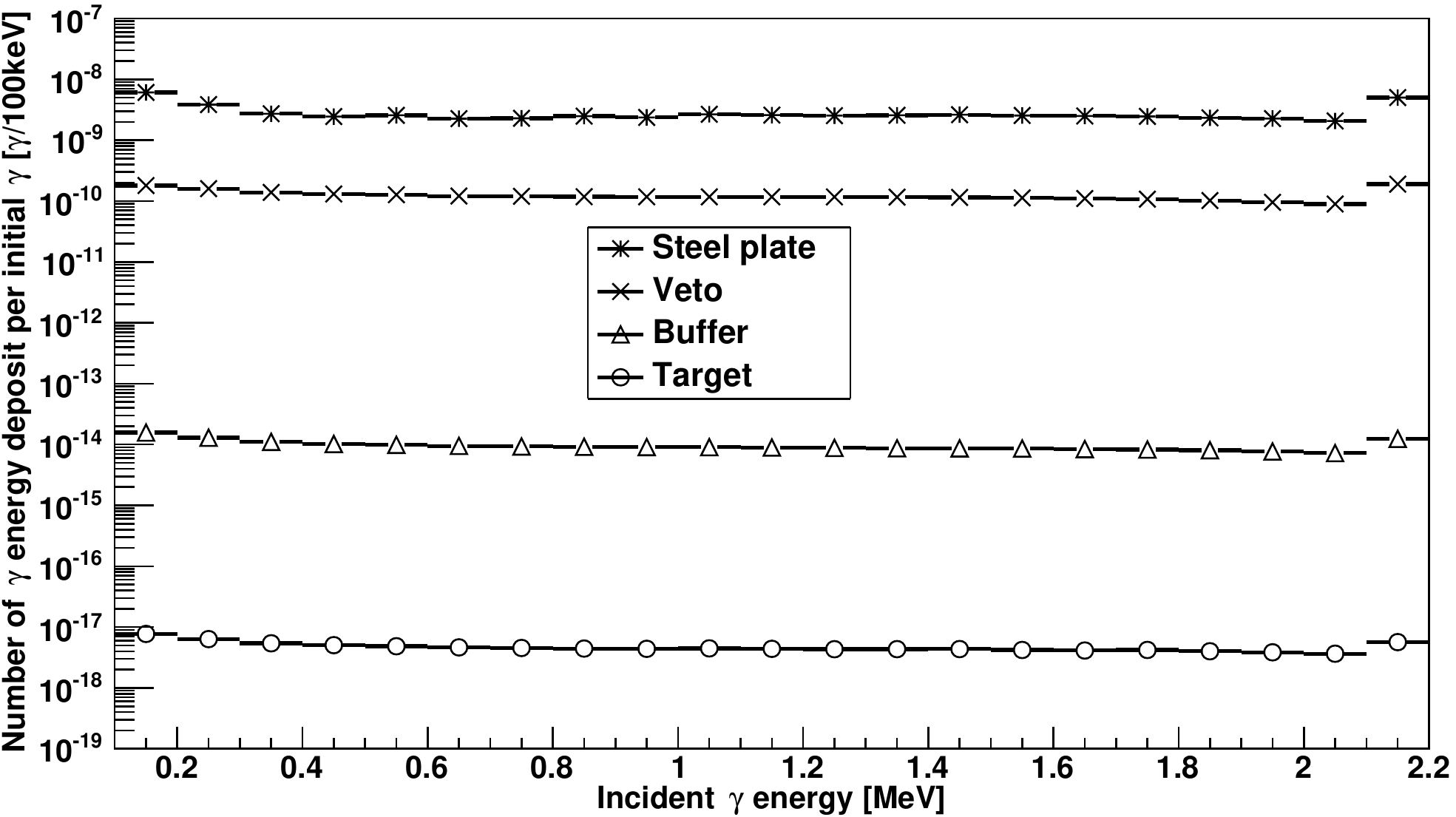}
\includegraphics[width=0.49\linewidth]{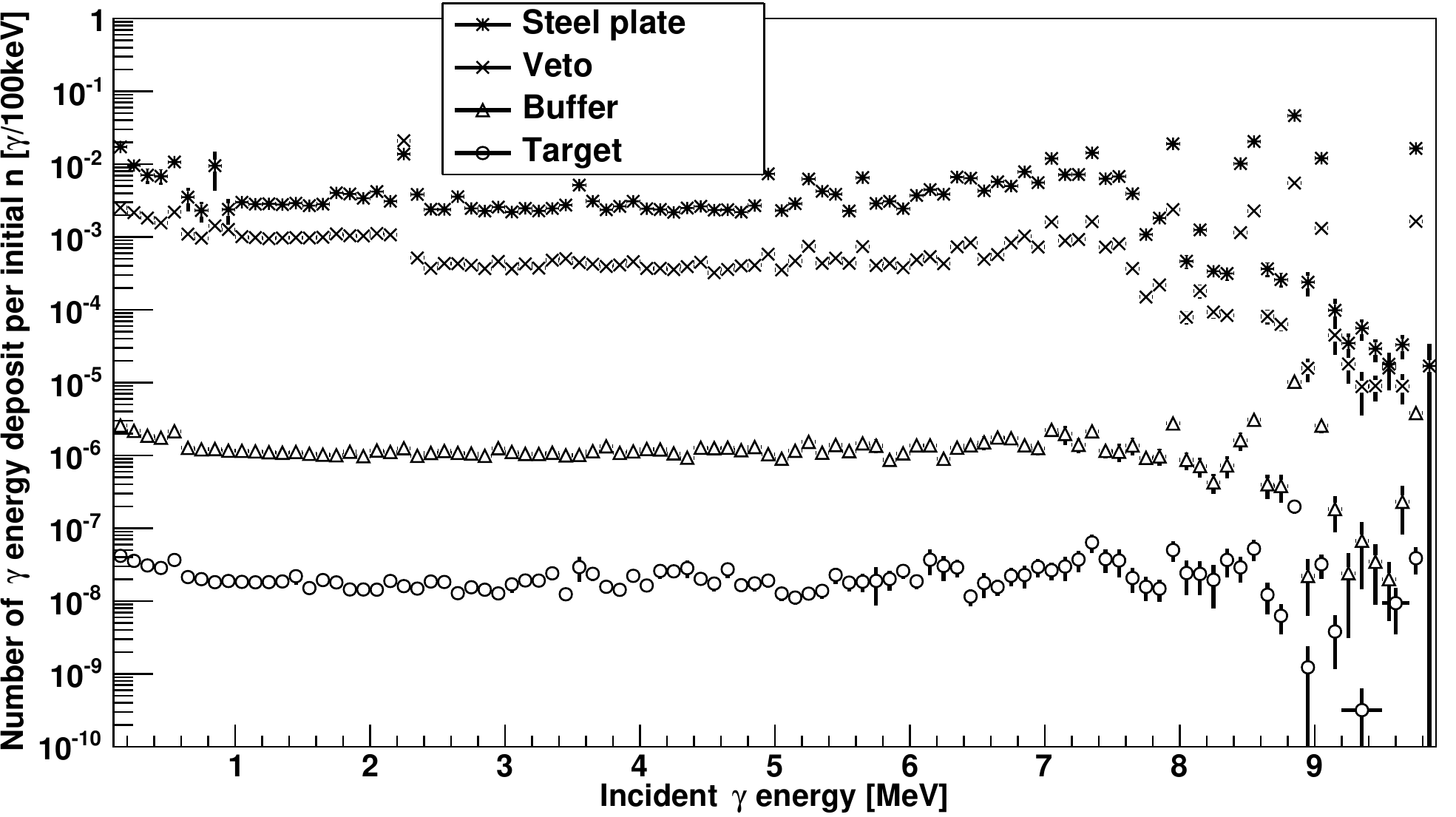}
\caption{\label{fig:attenuation} Simulation of the number of gamma energy deposits from the steel plate to the
target, per one initial gamma ray at \SI{2.185}{MeV} (left) and per one initial neutron shot in a
Watt spectrum (right). Errors bars are statistical only. The gamma simulation being much faster and easier to
bias, the statistics is higher and errors bars are contained in the points. The neutron simulation shows
several points with very high errors bars, a symptom of poor convergence in the considered bins. It is worth
noting that direct neutron energy deposits are not considered here, only the energy deposits of gammas from
neutron reactions. High energy gammas mainly come from neutron capture on iron, as indicated by the peaks in
the spectrum in steel plate. At lower energy, the \SI{2.2}{MeV} peak signs the neutron capture on hydrogen
and the \SI{0.5}{MeV} peak comes from pair creation.
}
\end{figure*}

Left panel of figure~\ref{fig:attenuation} shows the probability of a gamma to interact in the
different media (steel plate, veto, buffer, target volume - see figure~\ref{fig:genericsetup}) along its path
to the target volume as a function of energy. For instance, \num{6e-17} gamma per initial \SI{2.185}{MeV}
gamma emitted by the source interact in the target with an incident energy in the range \SIrange{1}{2.4}{MeV}
(the typical prompt energy window for the selection of IBD candidates). In the delayed energy window, only
\num{9.3e-18} gamma per initial  \SI{2.185}{MeV} gamma make an energy deposition in the target volume. The
total current of gamma ray with energies greater than \SI{1}{MeV} entering the target volume with an energy
higher than \SI{1}{MeV} amounts to \num{6e-17} $\upgamma$ per initial $\upgamma$, and is consistent with the
previous numbers. Hence, the expected count rate of events above \SI{1}{MeV} from source induced gamma ray is
\SI{1.3e2}{events/day} for a \SI{3.7}{PBq} source. Therefore, assuming a \SI{1}{ms} time coincidence window
and no delayed energy cut, a rate of \num{2e-4} accidental IBD-like events per day is expected (\SI{1}{event}
every \SI{13}{years}). Applying a delayed energy cut at \SI{>2}{MeV} reduces the accidental rate to \num{3e-5}
IBD-like events per day.

\begin{table}[!ht]
 \centering
 \begin{tabular}{c >{\centering}m{1.5cm} >{\centering}m{1.3cm} >{\centering}m{1.8cm} c}
  \toprule
  	\rule{0pt}{2.25ex}
  Material & Thickness (cm) & Density (\si{g/cm^3}) & Att. coeff. (\si{cm^2/g})
    & Tot. Att.  \tabularnewline \colrule
  	\rule{0pt}{2.5ex}
  $\text{CeO}_2$ & \num{5.58} & \num{4.5} & \num{4.06e-2} & \num{3.6e-1} \tabularnewline
  W alloy	& \num{19}  & \num{18.5} & \num{4.26e-2} & \num{3.1e-7} \tabularnewline
  Steel		& \num{10}  & \num{7.87} & \num{4.10e-2} & \num{4.0e-2}\tabularnewline
  Water		& \num{200} & \num{1.0}  & \num{4.69e-2} & \num{8.5e-5} \tabularnewline
  Oil		& \num{200} & \num{0.77} & \num{4.50e-2} & \num{9.8e-4} \tabularnewline
  \botrule
 \end{tabular}
 \caption{\label{tab:attenuation} Main materials contributing
to gamma attenuation in our generic reference experiment. The total
attenuation coefficient is taken from NIST's XCOM database~\cite{NIST-XCOM}.
The $\text{CeO}_2$ thickness is the effective thickness as described in
equation~\ref{eq:eff_thick}. }
\end{table}

These results can be worthily compared to an analytical calculation of the attenuation $\mathtt{A}$ of a
gamma ray propagating along the z-axis from the source to the target:
\begin{equation}
\mathtt{A} = \exp \left(- \sum_k e_k \ \lambda_k \ \rho_k \right)
\end{equation}
where $e_k$, $\rho_k$ and $\lambda_k$ are respectively the thickness, the density and the
total attenuation coefficient of the $k^{th}$ medium being crossed by the gamma-ray from the source to the
detection volume. An averaged material thickness must be considered for the $\text{CeO}_2$ source material, 
because gamma rays can be emitted anywhere along the height of the source. The $\text{CeO}_2$ source material
thickness is calculated according to equation~\ref{eq:eff_thick}:
\begin{equation}
 \label{eq:eff_thick}
 \exp \left( - \rho \ \lambda \ e_{\text{eff}} \right) = \frac{1}{H} \int_0^H
e^{-x \rho \lambda} \ud x
\end{equation}
with $H=\SI{14}{cm}$ the height of the source, which gives $e_{\text{eff}} = \SI{5.58}{cm}$.

Considering the media listed in table~\ref{tab:attenuation} along with their corresponding thickness,  density
and attenuation coefficient, the total attenuation factor is found to be \num{3.7e-16} at \SI{2.185}{MeV}. The
previous calculation must be refined by taking into account the source-detector solid angle. The solid angle
can be calculated under the approximation of a point-like source with equation~\ref{eq:solid_angle}:
\begin{equation}
 \label{eq:solid_angle}
 \frac{\Omega}{4 \pi} = \frac{1}{2} \left( 1-\sqrt{1-\left(
\frac{R}{L}\right)^2} \right)
\end{equation}
where $R$ is the radius of the detection volume and $L$ is the distance between the
center of the source and the center of the target volume. For the generic reference experiment,  $\Omega/4\pi
= \SI{8.24}{\%}$, so that the total attenuation is \num{3e-17}. This has to be compared with the \SI{6e-18}{}
attenuation factor in \SIrange{2.1}{2.2}{MeV} energy range obtained from the simulation.

This calculation overestimates the number of \SI{2.2}{MeV} gamma rays reaching the target volume by a factor
5. Actually, this calculation assumes that all gamma-rays propagate along the z-axis and therefore with a
minimal path length (i.e $\text{CeO}_2$ source thickness), while the simulation considers any paths from the
source to the target volume.

Another refinement of the analytic modeling considers a coupled calculation of the solid angle and attenuation
factor. The angle $\mathrm{\theta}$ between the z-axis and the baseline defined by the source center and
gamma interaction point is only considered here (i.e. the source is assumed to be spherical). As shown by
equation~\ref{eq:cos}, each material thickness is then increased by a factor $1/\cos \theta$:
\begin{align}
\label{eq:cos}
\mathtt{A} &= \frac{1}{4\pi} \int_{-\pi}^\pi \ud \varphi \int_0^{\theta_{max}} \exp \left(
-\sum_k \frac{e_k}{\cos \theta} \lambda_k \ \rho_k \right) \sin \theta \ud \theta
\end{align}
where $\mathrm{\theta_{max}}$ is the maximum $\mathrm{\theta}$ angle subtended by the detection volume
and is defined by $\mathrm{\sin \theta_{max} = R/L}$. Computing numerically this equation for the generic
experiment configuration leads to an attenuation factor of \SI{5e-18}, similar to the simulation results. This
simple model therefore brings useful information and allows to correctly estimate the rate of source induced
gamma backgrounds in any experimental setup, without the need of a dedicated simulation package.

\subsection{Source induced neutron background}
Actinide contaminants require a great attention, since in addition to $\upgamma$ and $\upbeta$ emission
they can undergo $\upalpha$ decays and spontaneous fission (SF). Emission of $\upbeta$ and $\upgamma$ rays
will be completely suppressed by the tungsten shielding, as explained in the previous section, whereas
$\upalpha$ particle emission will be absorbed by the source itself. Actinides emits fast neutrons through
$\upalpha$ interaction with light nuclei such as oxygen in the $\text{CeO}_2$ or spontaneous fission. The
latter process is particularly dominant for neutron rich heavy nuclei with an even number of nucleons. Those
fast neutrons can easily escape the source despite the high-Z shielding and scatter out in the detector
materials to finally produce gamma rays outside the shielding through radiative capture. Such captures mostly
occur on hydrogen atoms in water and oils (such as non-doped scintillators), producing \SI{2.2}{MeV}
$\upgamma$ rays which can mimic both the prompt and the delayed events of an IBD reaction. As opposed to the
\SI{2.185}{MeV} $\upgamma$ rays that can be attenuated through the high-Z shielding, the neutron emission
could be the dominant source of induced backgrounds events even though the actinide contamination level is
small.

However, the specific neutron activity of each of the actinide contaminants has to be balanced by their
respective abundance in spent nuclear fuel. Fresh nuclear fuel does not contain any isotope heavier than
\uh{}. All heavy isotopes are produced through a long chain of neutron capture and $\upbeta^{-}$ decay.
Moreover, nuclei with odd number of nucleons generally have large fission cross-sections, counterbalancing the
radiative capture chain and the production of heavier isotopes. A burn-up threshold is therefore expected,
depending on the isotope mass, together with a constant ratio between heavy isotope quantities, depending on
cross-sections, neutron fluxes and decays. Therefore, minor actinides are much less produced than lanthanides
during irradiation. Furthermore, the heavier the actinide, the lower is its abundance in spent nuclear fuel.
This effect will compensate the increase of branching ratio to SF with the number of nucleons.

\begin{table}[!ht]
 \centering
 \begin{tabular}{c c c >{\centering}m{2cm} >{\centering}m{2.1cm}}
  \toprule
  	\rule{0pt}{2.25ex}
  	Isotope & Half-life & $I_{\text{SF}}$ (\si{\%}) & Specific neutron activity
(\si{n/g}) & $\mathcal{A}$ / $\mathcal{A}$(\ce{}) in~SNF (Bq/Bq)\tabularnewline \colrule
 	\rule{0pt}{2.5ex}
$^{241}\text{Am}$	& \SI{432.2}{y}	& \num{4.0e-10}	& \num{1.2}	&\num{5e-03}	\tabularnewline
$^{242m}\text{Am}$	& \SI{141}{y}	& \num{4.7e-9}	& \num{46}	&\num{1e-04}	\tabularnewline
$^{243}\text{Am}$	& \SI{7370}{y}	& \num{3.7e-9}	& \num{0.72}	&\num{2e-04}	\tabularnewline
\tabularnewline
$^{240}\text{Cm}$	& \SI{27}{d}	& \num{3.9e-6}	& \num{6.7e7}	&		\tabularnewline
$^{242}\text{Cm}$	& \SI{162.8}{d}	& \num{6.2e-6}	& \num{1.9e7}	&		\tabularnewline
$^{243}\text{Cm}$	& \SI{29.1}{y}	& \num{5.3e-9}	& \num{2.6e2}	& \num{2e-04}	\tabularnewline
$^{244}\text{Cm}$	& \SI{18.10}{y}	& \num{1.4e-4}	& \num{1.6e7}	& \num{2e-02}	\tabularnewline
$^{245}\text{Cm}$	& \SI{8.5e3}{y}	& \num{6.1e-7}	& \num{1.1e2}	& \num{3e-06}	\tabularnewline
$^{246}\text{Cm}$	& \SI{4.73e3}{y}& \num{3.0e-2}	& \num{1.0e7}	& \num{5e-07}	\tabularnewline
$^{248}\text{Cm}$	& \SI{3.40e5}{y}& \num{8.39}	& \num{4.2e7}	&$<$\num{1e-07}	\tabularnewline
\tabularnewline
  \botrule
 \end{tabular}
 \caption{\label{tab:neutron_activity} Neutron emission and activity $\mathcal{A}$ relative to \ce{} activity
of most produced Am and Cm isotopes~\cite{NNDC}.}
\end{table}

Table \ref{tab:neutron_activity} shows the half-life, branching ratio to SF and specific neutron activity for
the most produced isotopes of americium and curium with half-life higher than \SI{180}{days}. 
It also provides a relative estimate of the activity of these minor actinides in 2-year old spent nuclear
fuel, normalized to the \ce{} activity.
The nuclei of interest are the even curium isotopes ($^{244,246,248}\text{Cm}$) because all americium isotopes
and odd nucleon number isotopes have very low branching ratio to SF ($\mathrm{\lesssim \SI{e-5}{\%}}$).
Berkelium isotopes have very short period, and nuclei heavier than $^{248}\text{Cm}$ ($^{250}\text{Cm}$, Cf,
Es, Fm) can be safely ignored because either their abundance in spent nuclear fuel is negligible or their
half-life period is short.
It is worth noting that the mean number of neutron released per SF $\upsilon$ increases with the nucleus mass,
following roughly $\upsilon=0.1094\:A - 23.94$ in the mass range of the previously discussed actinide
contaminants, so that $^{248}\text{Cm}$ releases \SI{33}{\%} more neutrons per SF than $^{241}\text{Am}$
($\upsilon$ = 3.2 and $\upsilon$ = 2.4 respectively).

Taking into account all these effects shows that the \cm{} is the most problematic
nucleus from the SF induced background point of view, with a \SI{18}{y} life-time,
a branching ratio to SF of \SI{1.4e-4}{\%} and a typical production of a few grams
per fuel assembly in standard VVER-400 cycles \cite{Kornoukhov:1997cm,Zaritskaya1981}. 
Roughly speaking, the expected production of isotopes heavier than \cm{} is lower by one 
order of magnitude per additional nucleon, leading to negligible quantities of Cf and heavy Cm isotopes.
All together, minor actinides (therefore excluding U, Pu and Np) will produce about
\SI{e7}{neutrons/g} after \SI{3}{years} of cooling, for typical VVER-400 spent
fuel~\cite{Mayak}. The actinides are efficiently separated from cerium during the ANG production, but 
remaining trace contaminations are expected.  Assuming \SI{e-5}{Bq} of \cm{} per Bq~of \ce{} we computed
 that \SI{1.4e5}{n/s}  are expected to be emitted by the ANG.

The simulation of source-induced neutron background in the generic reference experiment is done by
randomly shooting single neutrons following a Watt fission spectrum, and distributed homogeneously in the
source. Only \SI{60}{\%} of neutrons are captured by the
shielding. Surviving neutrons are captured in the ground, in the steel plate or in the
veto. Captures on metal produce high energy gamma rays (\SIrange{6}{9}{MeV}), while
captures on hydrogen in the veto produces \SI{2.2}{MeV} gamma rays. All of these gamma
rays are then degraded on their travel to the target volume, which tend to flatten the capture peaks.

Right panel of figure~\ref{fig:attenuation} shows the probability of making a gamma energy deposition in the
different detector sub-volumes consequently to the emission of one neutron, as obtained with the
TRIPOLI-4\textregistered{} software. For instance, the probability of making a gamma energy deposition in the
\SIrange{1}{2.4}{MeV} prompt energy range is \SI{2.4e-7}{}, with an almost flat spectrum in this range. In the
\SIrange{2.0}{2.4}{MeV} delayed energy range, the corresponding probability is \SI{0.6e-7}{}. A contamination
of \SI{e-5}{Bq} of \cm{} per Bq~of \ce{}  would lead to \num{0.1} and \num{\sim0.02} accidental IBD-like
events per day without and with a \SI{>2}{MeV} energy cut for the delayed event selection, respectively. The
neutron induced accidental background would then still be negligible as compared to the expected \anu{} rate,
although it is higher than the gamma induced accidental background by 3 order of magnitudes.

\begin{table}[!ht]
 \centering
 \begin{tabular}{c c c c }
  \toprule
 	            & $\anu{}$ & $\upgamma$ induced& neutron induced \tabularnewline
 	Activity &   (/day)          &  IBD-like signal &   IBD-like signal 
\tabularnewline \colrule
 	\rule{0pt}{2.5ex}
  \SI{5.5}{PBq}   & 80        &  \num{6.5e-5}               & \num{3.6e-2}   \tabularnewline
  \SI{3.7}{PBq}   & 54        &  \num{3.0e-5}               & \num{1.6e-2}   \tabularnewline
  \SI{2.7}{PBq}   & 39        &  \num{1.5e-5}               & \num{0.1e-2}   \tabularnewline
  \SI{1.7}{PBq}   & 25        &  \num{0.6e-5}               & \num{0.3e-2}   \tabularnewline
  \SI{0.9}{PBq}   & 13        &  \num{0.2e-5}               & \num{0.1e-2}   \tabularnewline
\tabularnewline
  \botrule
 \end{tabular}
 \caption{\label{tab:signal_bkg} Summary of the expected $\anu{}$ daily signal and ANG induced background
rates for the generic experimental configuration. A contamination of \SI{e-5}{Bq} of \cm{} per
Bq~of \ce{} is assumed. IBD-like events are selected as follows: E$_{\rm prompt}$ in \SIrange{1}{2.4}{MeV}, 
E$_{\rm delayed}$ in \SIrange{2.0}{2.4}{MeV} and $\Delta t_{pd}<$ \SI{1}{ms}. Backgrounds are varying
quadratically with the activity.}
\end{table}

Fast neutron emission from actinide contaminants in the source could also in principle be a source of
correlated backgrounds. The first type of correlated background arise from single fast neutrons reaching the
detection volume. Such neutrons can make a proton recoil and fake a prompt signal. Then, the neutron capture
would complete the coincidence signal with a correlation time similar to IBD, the latter being driven
by neutron thermalization and diffusion. However, neutrons have a very small probability to cross the inactive
regions of the detector.
Despite the use of the exponential biasing techniques in TRIPOLI-4\textregistered{}, no neutrons reached  the
detector target in our simulations, and only few of them reached the buffer, leading to an upper limit of
\num{1e-12} neutrons reaching the buffer per initial neutron.
Furthermore, only the most energetic neutrons reaching the liquid scintillator volume
would be able to make an energy deposition larger than the IBD \SI{1}{MeV} visible energy threshold. The
probability of such an energy deposition to be observed is even further reduced because of the high quenching
factor of protons in liquid scintillators.

The second type of potential correlated background follows from multiple neutron emission in actinide SF. Two
neutrons could scatter out of the source in the detector direction and be captured on hydrogen inside or close
to the detection volume. The $\upgamma$ rays from both captures would finally fake the IBD signal, the time
correlation being ensured by the capture time of the two neutrons.
Since two neutron-induced energy depositions are required to mimic a IBD signal, this background scales
quadratically with the energy deposition in the scintillator per initial neutron. The
TRIPOLI-4\textregistered{} simulation of neutrons for the generic reference experiment was unable to reproduce
such correlated events in the target volume. Interpolating the results obtained from a simulation of an
experimental configuration with thinner veto and buffer volumes, an upper limit of \num{e-4} IBD-like events
per day could be derived.
Finally, the combination of a \cm{} contamination of less than ${\cal{O}}(\SI{e-4}{Bq/Bq})$
combined with the thick veto and buffer detector volumes would lead to negligible source-induced correlated
backgrounds.

In any deployment scenario, neutron background could be further reduced by adding a neutron shielding, made of
a moderator and a neutron absorber. Typical neutron moderators are neutron-rich materials, such as mineral
oils, water or plastics. Using boron is the common solution to absorb neutrons without gamma emission, because
the $^{10}\text{B}(n,\upalpha)$ reaction has a very high cross-section and releases only a \SI{478}{keV} gamma
ray. Outside the detector, the ANG would typically be shielded with borated polyethylene or polyethylene
coated by boron carbide $\text{B}_4\text{C}$. With an ANG at the center of a detector, neutron backgrounds (if
any) would tremendously complicate or even prevent a source experiment such as described here. A dedicated
neutron shielding surrounding the gamma shielding would be mandatory, such as a balloon filled with saturated
boric water. It is worth noting that even with such a configuration, about $1/10^4$ neutrons are captured on
hydrogen, since the ratio of thermal radiative capture cross sections is~\cite{NNDC}:
\begin{equation}
\frac{\sigma_{^{10}\text{B}(n,\upalpha)}}{\sigma_{\text{H}(n,\upgamma)}} =
\frac{\SI{3803}{b}}{\SI{0,3284}{b}} = \num{1.16e4}
\end{equation}
This ratio could be increased by an other factor 656 using heavy water, since deuterium
thermal radiative capture cross-section is only \SI{5.0e-4}{b}~\cite{NNDC}.

\section{Conclusion}
A definitive test of the short baseline anomalies is necessary to address the hypothesis of a light sterile
neutrino. A smoking gun signature of neutrino oscillations at short distances is the observation of an
oscillation pattern both in the reconstructed energy spectrum and spatial distribution of the IBD events. Such
an observation can be performed in a short-term and at a (relatively) modest cost through the deployment of an
intense ANG near an existing large liquid scintillator detector. The \ce{}-\pr{} couple has been identified to
be the most suitable \anu{} emitter for this type of experiment. The production of a \ce{}-\pr{} ANG with a
few PBq activity is technically feasible, and can be realized by the Russian \ma{} company by reprocessing
spent nuclear fuel. Detectors with ultra low background levels and good vertex and energy reconstruction
capabilities are necessary for the observation of an oscillation pattern in the $(\Dmn,\,\stn)$ parameter
space relevant to the neutrino anomalies at short baselines. Three large liquid scintillator detectors which
meet these requirements have been identified and are namely, the KamLAND, Borexino and SNO+ detectors.
Deploying an intense ANG next to one of these detectors allows to completely neglect any kind of
detector-induced backgrounds. As pointed out in this article, another possible source of backgrounds comes
from the ANG itself. It has been shown that the impact of the source-induced backgrounds can be severely
limited provided that the ANG is enclosed in a thick $\sim$\SI{20}{cm} high density material shielding (such
as tungsten) and that the neutron emitting radioactive contaminants, which are leftovers resulting from the
source manufacturing process, are kept at sufficiently low levels.

More precisely, the activity of any $\upgamma$ emitters in the source must be smaller than \SI{1}{\%} relative
to the ANG absolute activity, such that the thickness of the high-density material shielding is driven by the
activity of the \SI{2.185}{MeV} deexcitation gamma ray following the $\upbeta^-$ decay of \pr{}. The
contamination of minor actinides must be less than \SI{e-5}{Bq} of actinide per Bq~of \ce{}. Such a
contamination level brings the rate of the neutron induced accidental background to the rate of the
$\upgamma$-induced accidental background. A detailed simulation of gamma and neutron particle transport from
the source up to the detection volume confirmed that the resulting backgrounds are negligible. Even though
slightly higher contamination levels were observed, the sensitivity and physics reach of the experiment would
not be degraded.

The ANG absolute activity is also another important parameter to control, especially for optimizing the
experiment sensitivity in the high $\Dmn \gtrsim \SI{10}{eV^2}$. As such, the amount of $\upbeta$ radioactive 
impurities in the source must then be kept at a level of \SI{0.1}{\%} or less to prevent any bias in the 
determination of the ANG activity with a calorimetric method. Moreover, the source $\upbeta$ and \anu{} 
spectrum shapes must be precisely measured in order not to spoil both the precision of the ANG absolute 
activity and the experiment sensitivity in a rate free analysis.

Having these source and detector requirements fulfilled ensures the sensitivity of such an experimental
configuration to be mostly driven by statistical uncertainties. The best experimental scenario then
corresponds to the biggest detector, the closest deployment location and the highest source activity. If any,
higher unexpected source induced backgrounds could be mitigated by a more distant deployment location and
additional high or/and low density material shieldings.

To assess the impact of a future ANG experiment for ruling out the electron dissapearance anomalies we combined
 the reactor and gallium results with the future measurements of each ANG experiment.  
We used the parameter goodness of fit (PGof) test based on the difference between the overall best fit $\chi^2$
 and the sum of the best fit $\chi^2$'s for each ANG experiment only and the combination of both the reactor 
and gallium anomalies~\cite{Maltoni:2003cu}. 
If any ANG experiment would report a no-oscillation result the PGof resulting probabilities would be down to the level of~1-1.5\%.
The allowed mixing angles would be restricted to $\sin^2(2\theta) \lesssim $0.05  for $\Delta m^2 \lesssim$ 1.5~eV$^2$.

However the sensitivity the ANG experiment could be improved by either increasing the source activity (if technologically 
feasible) or by deploying the source closer to the active target volume or even inside the detector. The 
latter cases would necessitate a refurbishment of the detector as well as a more demanding control of 
the source induced backgrounds. In particular the mitigation of the neutron induced 
backgrounds would need stringent requirements on the minor actinides contaminants, probablyh leading to further 
purification steps during the production.

In case of hint of any oscillation signal another option would be to relocate the ANG at a slightly different baseline 
during the course of the experiment. The comparison of data at two different distances would be a reliable test against any oscillation scenario

\section*{Acknowledgement}
We thanks the KamLAND collaboration for their active work and full support towards the investigation of the
CeLAND experimental configuration. We acknowledge the continuous strong support of the Borexino collaboration
towards the realization of the CeSOX experiment at the Laboratori Nazionali del Gran Sasso. We are grateful to
Russian \ma{} company for their research and developments towards the realization of an intense \ce{} ANG.
Finally we would like to thank V.~Kornoukhov for its continuous efforts on the engineering of the \ce{} ANG.
Th.~Lasserre thanks the European Research Council for support
under the Starting Grant StG-307184.

\bibliography{2015-02-09_parameters-CeANG_bibliography}

\end{document}